\documentclass[12pt]{article}
\usepackage{amsmath,amssymb}
\usepackage{graphicx}
\begin{document}

\title{Dynamic Depletion of Vortex Stretching and Non-Blowup of the 3-D 
Incompressible Euler Equations}
\author{Thomas Y. Hou\thanks{Applied and Comput. Math, 217-50, Caltech, Pasadena, 
CA 91125. Email: hou@acm.caltech.edu, and LSEC, 
Academy of Mathematics and Systems Sciences, Chinese Academy of Sciences,
Beijing 100080, China.}
\and Ruo Li\thanks{Applied and Comput. Math., Caltech, Pasadena, CA 91125, and
LMAM\&School of Mathematical Sciences, Peking University, Beijing, China.
Email: rli@acm.caltech.edu.}}

\maketitle

\begin{abstract}
We study the interplay between the local geometric properties
and the non-blowup of the 3D incompressible Euler equations. 
We consider the interaction of 
two perturbed antiparallel vortex tubes using Kerr's initial condition
\cite{Kerr93}[Phys. Fluids {\bf 5} (1993), 1725]. We use a 
pseudo-spectral method with resolution
up to $1536\times 1024 \times 3072$ to resolve the nearly 
singular behavior of the Euler equations. Our numerical results 
demonstrate that the maximum vorticity does not grow 
faster than double exponential in time, up to $t=19$, beyond the 
singularity time $t=18.7$ predicted by Kerr's computations 
\cite{Kerr93,Kerr04}. The velocity, the enstrophy and
enstrophy production rate
remain bounded throughout the computations. As the flow evolves, 
the vortex tubes are flattened severely and turned into thin 
vortex sheets, which roll up subsequently. The vortex lines near 
the region of the maximum vorticity are relatively straight. 
This local geometric regularity of vortex lines seems to be 
responsible for the dynamic depletion of vortex stretching.
\end{abstract}

\section{Introduction}

One of the most challenging questions in fluid dynamics is whether the 
incompressible 3D Euler equations can develop a finite time singularity
from smooth and bounded initial data. From a theoretical point of view, the 
main difficulty is due to the presence of the vortex stretching term in 
the vorticity equation, which is formally quadratic in vorticity. If 
such quadratic nonlinearity persists in time long enough, we would expect  
a finite time singularity of the form $O(T-t)^{-1}$ in vorticity.
Such blow-up rate is consistent with the well-known result of Beale-Kato-Majda
\cite{BKM84} (see also \cite{EFM70}). There have been many computational 
efforts in searching for
finite time singularities of the 3D Euler and Navier-Stokes equations, see e.g. 
\cite{Chorin82,PS90,KH89,GS91,SMO93,Kerr93,Caf93,BP94,Pelz98,GMG98,Kerr04,Chorin06}. 
One example that has been studied extensively is the interaction of two 
perturbed antiparallel vortex tubes. This example is interesting because of 
the vortex reconnection which has been observed for the corresponding 
Navier-Stokes equations. It is natural to ask whether the 3D Euler 
equations would develop a finite time singularity in the limit of 
vanishing viscosity. 

In \cite{Kerr93}, Kerr presented numerical evidence which suggests 
a finite time singularity of the 3D Euler equations for two perturbed 
antiparallel vortex tubes. In Kerr's computations, he used a 
pseudo-spectral discretization in the $x$ and $y$ directions, and
a Chebyshev discretization in the $z$ direction with resolution of order
$512\times 256 \times 192$. His computations showed that the growth of 
the peak vorticity, the peak axial strain, and the enstrophy production 
obey $(T-t)^{-1}$ with $T = 18.9$. Self-similar development and
equal rates of collapse in all three directions were shown
(see the abstract of \cite{Kerr93}). While
velocity blowup was not documented in \cite{Kerr93}, Kerr showed
in his subsequent papers \cite{Kerr97,Kerr99,Kerr04} that velocity
field blows up like $O(T-t)^{-1/2}$ with $T$ being revised to $T=18.7$. 
Kerr's computations have generated a lot of interests and his proposed
initial conditions have been 
considered as ``the most attractive candidates for potential singular 
behavior'' of the 3D Euler equations (see page 187 of \cite{MB02}). 

Vortex reconnection of two perturbed antiparallel vortex tubes has been 
studied extensively in the literature. Substantial core deformation has been
observed \cite{PS90,AG89,BPZ92,KH89,MH89,SMO93}. Most studies indicated 
only exponential growth in the maximum vorticity.  
However, the work of Kerr and Hussain in \cite{KH89} suggested a finite 
time blow-up in the infinite Reynolds number limit, which motivated 
Kerr's Euler computations mentioned above.

There has been some interesting development in the theoretical understanding
of the 3D incompressible Euler equations. It has been shown that the local 
geometric regularity of vortex lines can play an important role in depleting 
nonlinear vortex stretching \cite{Const94,CFM96,DHY05a,DHY05b}. In particular, 
the recent results obtained by Deng, Hou, and Yu \cite{DHY05a,DHY05b} 
show that geometric regularity of vortex lines, even in an extremely 
localized region containing the maximum vorticity, can lead to depletion of 
nonlinear vortex stretching, thus avoiding finite time singularity formation 
of the 3D Euler equations. To obtain these results, Deng-Hou-Yu 
\cite{DHY05a,DHY05b} explored the connection between the stretching 
of local vortex lines and the growth of vorticity. In particular,
they showed that if the vortex lines near the region of maximum vorticity 
satisfy some local geometric regularity conditions and the maximum velocity 
field is integrable in time, then no finite time blow-up is possible. See 
Section 4.2 for the detailed description of these results. Kerr's 
computations fall in the critical case of the non-blowup theory in
\cite{DHY05a,DHY05b}. To get a definite answer in this critical case, 
we need to check whether certain scaling constants, which describe 
the local geometric properties of the vortex lines, satisfy an 
algebraic inequality. However, such scaling constants are not 
available in \cite{Kerr93}. This is our original motivation
to repeat Kerr's computations.  

It is worth mentioning that the predicted singularity time in Kerr's 
computations is $T=18.7$, while his computations from $t=17$ to 
$t=18$,  as mentioned in \cite{Kerr93}, seem to be under-resolved and 
were ``not part of the primary evidence for a singularity''. 
Clearly, the computations for $t \le 17$, which Kerr used as the
primary evidence for a singularity, is still far from the predicted 
singularity time, $T=18.7$. In order to justify the predicted asymptotic 
behavior of vorticity and velocity blowup, one needs to perform 
well-resolved computations much closer to the predicted singularity 
time. Such well-resolved computations can also provide more accurate 
geometric properties of vortex lines, which can be used to check whether 
the non-blowup conditions in \cite{DHY05a,DHY05b} are satisfied.

In this paper, we perform well-resolved computations of the 3D incompressible
Euler equations using the same initial condition as the one used by Kerr in 
\cite{Kerr93}. A pseudo-spectral method with a very high order
Fourier smoothing is used to discretise the 3D incompressible Euler 
equations in all three directions. The time integration is performed using 
the classical fourth order Runge-Kutta method with adaptive time stepping to 
satisfy the CFL stability condition. We perform a careful numerical study to 
show that the pseudo-spectral method we use provides more accurate 
approximations to the 3D Euler equations than the pseudo-spectral method
that uses the standard 2/3 dealiasing rule. We use up to 
$1536\times 1024\times 3072$ space resolution in order to resolve 
the nearly singular behavior of the 3D Euler equations. 

Our numerical results demonstrate that the maximum vorticity does 
not grow faster than double exponential in time, up to $t=19$, beyond the
singularity time $t=18.7$ predicted by Kerr's computations \cite{Kerr93,Kerr04}.
There are three distinguished stages of vorticity growth in time. In the early 
stage for $0\le t \le 12$, the maximum vorticity grows only exponentially in time. 
During the intermediate stage for $12 \le t\le 17$, the two vortex tubes 
experience tremendous core deformation and become severely flattened. Each 
vortex tube effectively turns into a vortex sheet with rapidly decreasing 
thickness. During this stage, the maximum vorticity grows slightly slower 
than double exponential in time. It is also interesting to examine the 
degree of nonlinearity in the vortex stretching term during this stage. 
An $O(T-t)^{-1}$ blowup rate in the maximum vorticity would imply that the 
nonlinearity in the vortex stretching term is quadratic. However, our numerical 
results show that the vortex stretching term, when projected to 
the unit vorticity vector, is bounded by 
$ \|\vec{\omega} \|_\infty \log (\|\vec{\omega} \|_\infty )$, where 
$\vec{\omega} $ is 
vorticity. It is easy to see that such upper bound on the vortex stretching
term implies that the maximum vorticity is bounded by double exponential in time. 
During the final stage for $17 \le t \le 19$, we observe that the growth of 
the maximum vorticity slows down considerably and deviates from double exponential 
growth, indicating that there is stronger cancellation taking place in the 
vortex stretching term.

We also find that the vortex lines near the region of the maximum vorticity are 
relatively straight and the vorticity vectors seem to be quite regular. This 
was also observed in \cite{Kerr93}. On the other hand, the inner region 
containing the maximum vorticity does not seem to shrink to zero at a 
rate of $(T-t)^{1/2}$, as predicted by Kerr's computations. Moreover, we 
find that the velocity field, the enstrophy, and enstrophy production rate
remain bounded throughout 
the computations. The fact that the velocity field remains bounded is 
significant. With velocity field being bounded, the result of Deng-Hou-Yu 
\cite{DHY05a} can be applied, which implies the non-blowup of the Euler 
equations up to $T=19$. The geometric regularity of the vortex lines near 
the inner region seems to play an important role in the dynamic depletion 
of vortex stretching \cite{DHY05a,DHY05b}.

We would like to stress the importance of sufficient resolution in
determining the nature of the nearly singular behavior of the 3D Euler 
equations. As demonstrated by our numerical computations, the 3D Euler 
equations have different growth rates in the maximum vorticity  
in different stages. A resolution without the proper level of refinement 
would not be able to capture the transition from one growth phase to another. 
In \cite{Kerr93}, the inverse of the maximum vorticity was shown to approach 
to zero almost linearly in time up to $t=17$. If this trend were to continue 
to hold, it would lead to the blowup of the maximum vorticity in the form 
of $O(T-t)^{-1}$. However, with increasing resolutions, we find that 
the curve corresponding to the inverse of maximum vorticity starts to turn 
away from zero starting at $t=17$. We also observe that the velocity field 
becomes saturated around this time. Incidentally, this is precisely the 
time when Kerr's computations began to lose resolution. At $t=17$, the 
thin vortex sheets have already started to roll up. After $t=17.5$, the 
vorticity in the rolled up region has developed large gradients in all 
three directions, with $z$ being the most singular direction and $y$ being 
the least singular direction. To resolve the large gradients in all
three directions, we allocate $3072$ grid points along the $z$ direction, 
$1536$ along the $x$ direction, and $1024$ along the $y$ direction. This 
level of resolution ensures that we have about 8 grid points across the 
most singular region in each direction toward the end of the computations. 

Kerr interpreted the roll-up of the vortex sheet as ``two vortex sheets 
meeting at an angle'' and argued that the formation of this angle may be
responsible for the finite time blowup of the Euler equations. Our 
computations indicate that the rollup region of the vortex sheet 
is still relatively smooth even during the final stage of the 
computations. Moreover, according to the results in 
\cite{DHY05a,DHY05b}, it is the curvature of the vortex lines and 
the divergence of the unit vorticity vector that contribute to the blow-up, 
not the curvature of the vortex sheet itself. Further, we observe that the 
vortex lines near the region of the maximum vorticity are relatively smooth. 
This geometric regularity leads to strong dynamic depletion of the nonlinear 
vortex stretching. 

There is reason to believe that if the current scenario persists,
there is no blowup of the 3D Euler equations for these data beyond $T=19$. 
In fact, during the final stage of the computations for $17.5 \le t \le 19$, 
the vortex lines near the region of the maximum vorticity remain smooth. 
Further, as the vortex sheet rolls up, we observe that the location of 
the maximum vorticity moves away from the dividing plane separating the 
two vortex tubes toward the rolled up portion of the vortex sheet, leading 
to a slower growth rate of maximum vorticity. 

The rest of this paper is organized as follows. We describe the set-up of 
the initial condition in Section 2 and describe our numerical method in 
Section 3. In Section 4, we describe our numerical results in detail and 
perform comparisons with the previous results obtained in 
\cite{Kerr93,Kerr04}. Some concluding remarks are made in Section 5.

\section{The Initial Condition}

The 3D incompressible Euler equations in the vorticity stream function
formulation are given as follows:
\begin{eqnarray}\label{3deuler}
\vec{\omega}_t+(\vec{u}\cdot\nabla) \vec{\omega} & = & \nabla 
\vec{u} \cdot \vec{\omega}, \\
- \bigtriangleup \vec{ \psi} &= & \vec{\omega}, \quad
\vec{u} = \nabla \times \vec{\psi},
\end{eqnarray} 
with initial condition $\vec{\omega}\mid_{t=0} =  \vec{\omega}_{0}$,
where $\vec{u}$ is velocity, $\vec{\omega}$ is vorticity, and 
$\vec{\psi}$ is stream function. Vorticity is related to velocity 
by $\vec{\omega} = \nabla \times \vec{u}$. The incompressibility
implies that 
\[
\nabla \cdot \vec{u} = \nabla \cdot \vec{\omega} = \nabla \cdot \vec{\psi} = 0.
\]
We consider periodic boundary 
conditions with period $2 \pi$ in all three directions. 

We study the interaction of two perturbed antiparallel vortex tubes using 
the same initial condition as that of Kerr (see Section III of \cite{Kerr93}). 
There are a few misprints in the analytic expression of the initial 
condition given in \cite{Kerr93}. In our computations, we use the corrected 
version of Kerr's initial condition by comparing with Kerr's Fortran 
subroutine which was kindly provided to us by him.  A list of corrections
to these misprints is given in the Appendix.

The initial condition is given by a pair of perturbed anti-parallel vortex 
tubes, which is expressed in terms of vorticity. The vorticity that
describes the vortex tube above the $x$-$y$ plane is of the form:
\begin{equation}
  \vec{\omega} = \omega ( r ) (\omega_x, \; \omega_y,\; \omega_z ) .
\end{equation}
The first step in setting up the initial condition is to define the 
profile $\omega (r)$. 
If $r \ge 1$, we set $\omega ( r ) = 0$. For $r < 1$, we define 
\begin{equation}
  \omega ( r ) = \exp [ f ( r ) ],
\end{equation}
with $f(r)$ given by
\begin{equation}
f( r ) = \frac{- r^2}{1 - r^2} + r^4 \left( 1 + r^2 + r^4 \right) .\label{f}
\end{equation}
The radius $r$ is centered around an initial vortex core trajectory $(X,Y,Z)$, 
and is defined by
\begin{equation}
  r = \left| \left( x, y, z \right) - \left( X, Y, Z \right) \right| / R,
  \hspace{2em} \mathrm{for} \;\; r \leqslant 1 .
\end{equation}
The initial vortex core trajectory $(X,Y,Z)$ is characterized by
\begin{equation}
  \left( X, Y, Z \right) = ( x ( s ),  y,  z ( s ) ),
   \label{trajectory}
\end{equation}
where $s$ is a function of $y$ and
\begin{equation}
  x ( s ) = x_0 + \delta_x \cos ( \pi s / L_x ), \label{x-s}
\end{equation}
\begin{equation}
  z ( s ) = z_0 + \delta_z \cos ( \pi s / L_z ). \label{z-s}
\end{equation}
To complete the definition of $\omega (r)$, we need to define 
$s$ as a function of $y$, which is given below: 
\begin{equation}
  s ( y ) = y_2 + L_y \delta_{y 1} \sin ( \pi y_2 / L_y ),
\end{equation}
and
\begin{equation}
  y_2 = y + L_y \delta_{y 2} \sin \left( \pi y / L_y \right) .
\end{equation}

The second step is to define the vorticity vector 
$(\omega_x,\omega_y,\omega_z)$, which is given as follows:
\begin{eqnarray}
  \omega_x & =& - \frac{\pi \delta_x}{L_x} \left[ 1 + \pi \delta_{y_2} \cos
  \left( \frac{\pi y}{L_y} \right) \right] \times \left[ 1 + \pi \delta_{y 1}
  \cos \left( \frac{\pi y_2}{L_y} \right) \right] \sin \left( \frac{\pi s ( y
  )}{L_x} \right) ,\label{w-x} \\
  \omega_y & = & 1, \label{w-y} \\
  \omega_z & = & - \frac{\pi \delta_z}{L_z} \left[ 1 + \pi \delta_{y_2} \cos
  \left( \frac{\pi y}{L_y} \right) \right] \times \left[ 1 + \pi \delta_{y 1}
  \cos \left( \frac{\pi y_2}{L_y} \right) \right] \sin \left( \frac{\pi s ( y
  )}{L_z} \right). \label{w-z}
\end{eqnarray}

We choose exactly the same parameters as in \cite{Kerr93}.
Specifically, we set $\delta_{y 1} = 0.5$, 
$\delta_{y 2} = 0.4$, $\delta_x = -1.6$, $\delta_z = 0$, $z_0 = 1.57$
and $R = 0.75$. The constant $x_0$ fixes the center of perturbation 
for the vortex tube along the $x$ direction. In our computations, we
set $x_0 = 0$. Moreover, we choose $L_x = L_y = 4 \pi$, and $L_z = 2 \pi$.

The third step is to rescale the initial profile defined above. According to 
\cite{Kerr93} (see the last paragraph on page 1728 of \cite{Kerr93}), 
we need to rescale the above initial vorticity profile by a constant factor 
so that the maximum vorticity in the $y$ direction is increased to 8.
With the choice of the above parameters, the maximum vorticity in the $y$ 
direction before rescaling is equal to 0.999766. Thus, the constant rescaling 
factor is equal to 8.001873. 

The final step in defining the initial condition is to filter the initial
vorticity profile. After rescaling the initial vorticity, we apply 
exactly the same Fourier filter as the one used in \cite{Kerr93}, 
i.e. $\exp \left[ - 0.05 \left( k_x^4 + k_y^4 + k_z^4 \right) \right]$, 
to the initial vorticity to smooth the rough edges. 

\begin{figure}
\begin{center}
\includegraphics[width=8cm]{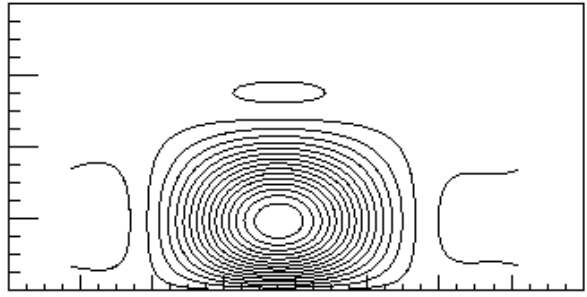}
\end{center}
\caption{The axial vorticity ($\omega_y$) contours of the initial value on 
the symmetry plane. \label{fig.vort-init}}
\end{figure}

\begin{figure}
\begin{center}
\includegraphics[width=8cm]{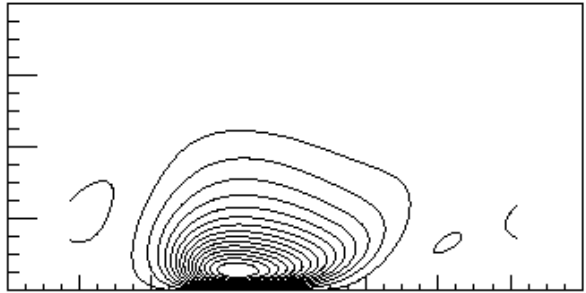}
\end{center}
\caption{The axial vorticity ($\omega_y$) contours when $t=6$ on 
the symmetry plane. \label{fig.vort-t=6}}
\end{figure}

\begin{figure}
\begin{center}
\includegraphics[width=8cm]{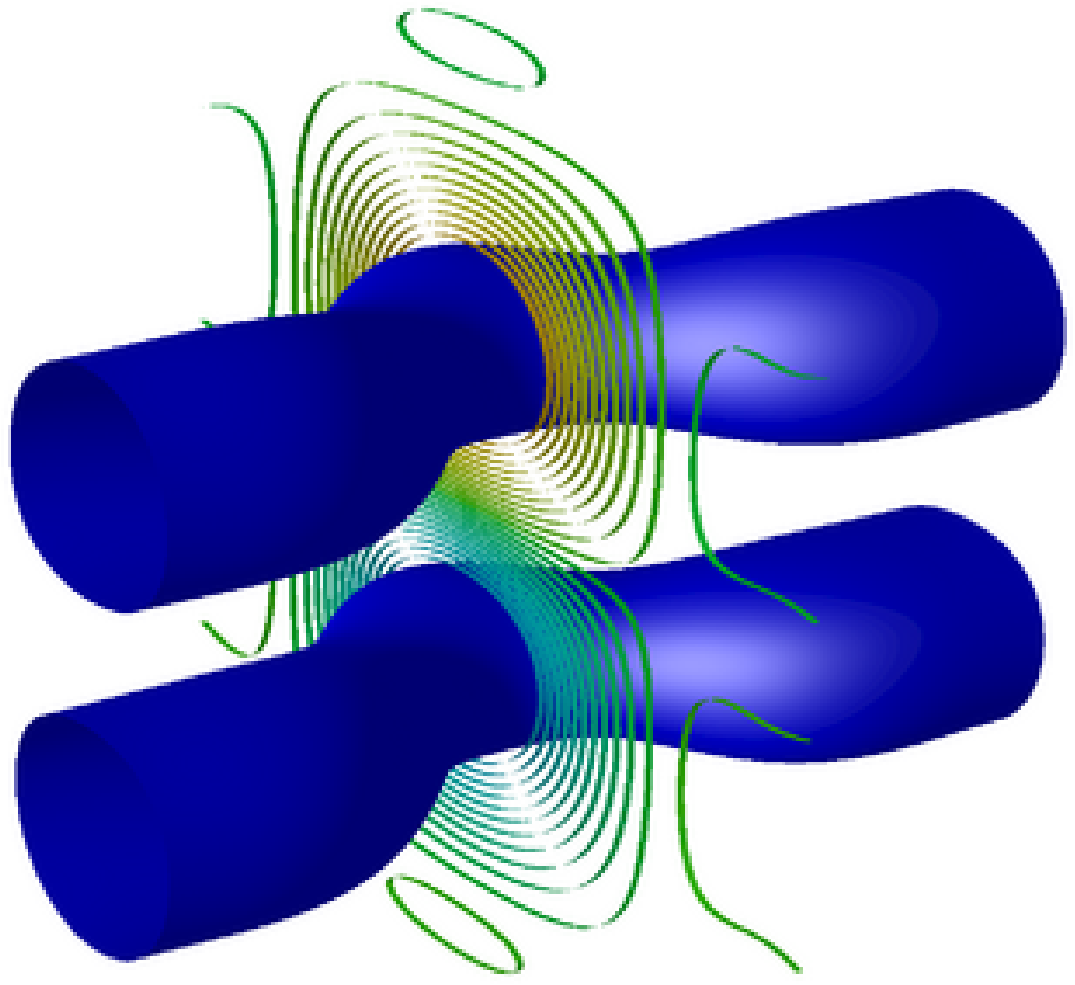}
\includegraphics[width=8cm]{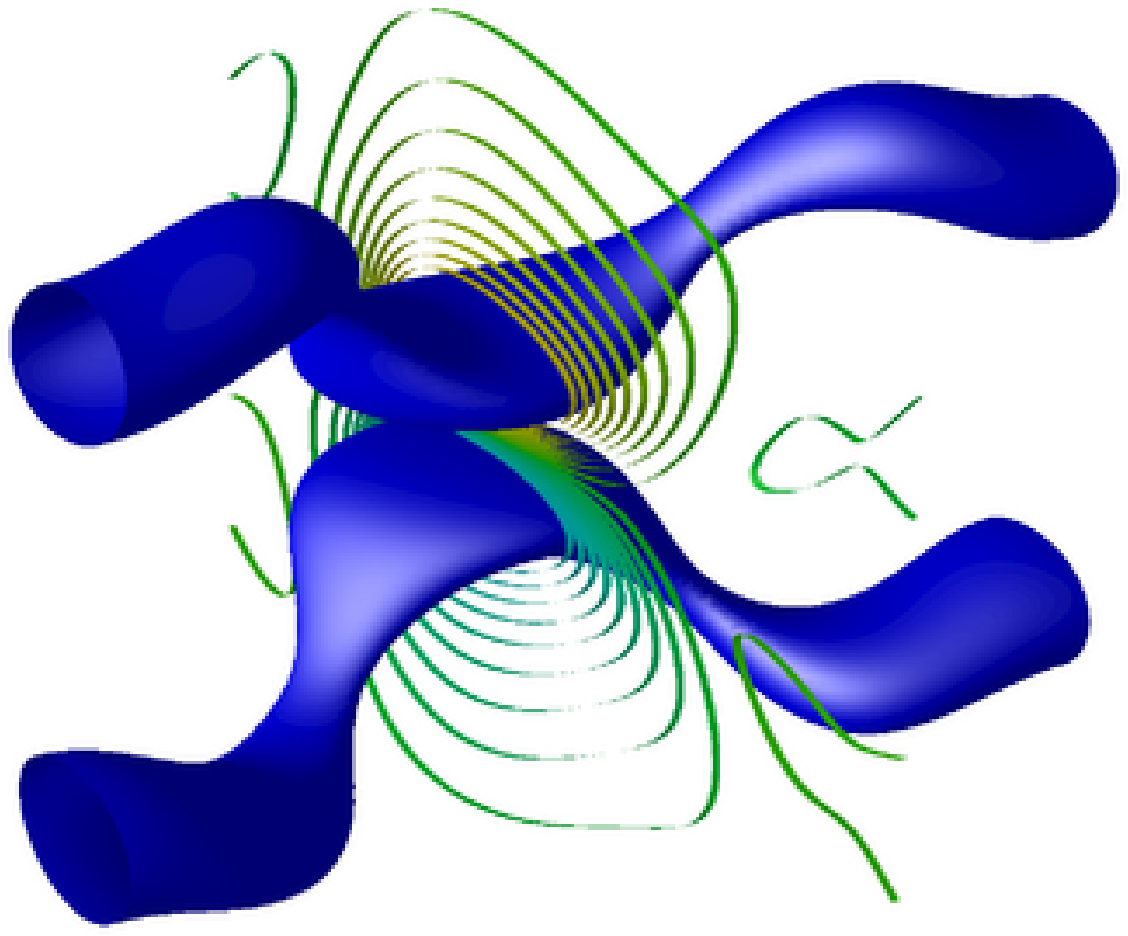}
\end{center}
\caption{The 3D view of the vortex tube for $t=0$ and $t=6$.
The tube is the isosurface at $60\%$ of the maximum vorticity.
The ribbons on the symmetry plane are the contours at other different
values. \label{fig.vorttube}} 
\end{figure}

We should point out that due to the difference between ours and Kerr's
discretization strategies in solving the 3D Euler equations, the discrete 
initial condition generated by Kerr's discretization and the one generated 
by our pseudo-spectral 
discretization are not exactly the same. In \cite{Kerr93}, Kerr used the
Chebyshev polynomials to approximate the solution along
the $z$ direction. In order to prepare the initial data that can be used for 
the Chebyshev polynomials, he 
performed some interpolation and used extra filtering. This interpolation
and extra filtering seem to introduce some asymmetry to Kerr's discrete
initial data. According to \cite{Kerr93} (see the top paragraph of 
page 1729), ``An effect of the initial filter upon the vorticity contours 
at $t=0$ is a long tail in Fig. 2(a)''. 
Since we perform pseudo-spectral approximations in all three directions, 
there is no need to perform interpolation or use extra filtering as was 
done in \cite{Kerr93}. To demonstrate this slight difference between 
Kerr's initial data and ours, we plot the initial vorticity contours 
along the symmetry plane in Figure \ref{fig.vort-init}. As we can see, 
the initial vorticity contours in Figure \ref{fig.vort-init} are 
essentially symmetric. This is in contrast to the apparent asymmetry
in Kerr's initial vorticity contours (see Fig. 2(a) of \cite{Kerr93}). 
The 3D plot of the initial vortex tubes is given in Figure \ref{fig.vorttube}. 
Again, we can see that the initial vortex tube is essentially symmetric. 
As time increases, the two antiparallel perturbed vortex tubes approach 
each other. By time $t=6$, we already observe a significant flattening 
near the center of the tubes, see Figure \ref{fig.vort-t=6} and 
Figure \ref{fig.vorttube}. 

\section{The Numerical Method}

We use the pseudo-spectral method with a very high order
Fourier smoothing to discretize the 3D Euler equations. 
The Fourier smoothing that we use along the $x_j$ direction is of the form: 
$\rho(2k_j/N_j) \equiv \exp(-\alpha (2k_j/N_j)^m)$ with $\alpha=36$ and 
$m=36$, where $k_j$ is the wave number ($|k_j| \leqslant N_j/2$) 
along the $x_j$ direction and $N_j$ is the total number of grid 
points along the $x_j$ direction. Specifically, if $\widehat{v}_k$
is the discrete Fourier transform of $v$, then we approximate
$v_{x_j}$ by taking the discrete inverse Fourier transform of 
$i k_j \rho(2k_j/N_j) \widehat{v}_k$, where $k =(k_1,k_2,k_3)$. 
The time integration is performed 
using the classical fourth order Runge-Kutta method. Adaptive
time stepping is used to satisfy the CFL stability condition with 
CFL number equal to $\pi/4$. We use up to 
$1536 \times 1024 \times 3072$ space resolution in order to resolve
the nearly singular behavior of the 3D Euler equations. 

\begin{figure}
\begin{center}
\includegraphics[width=8cm]{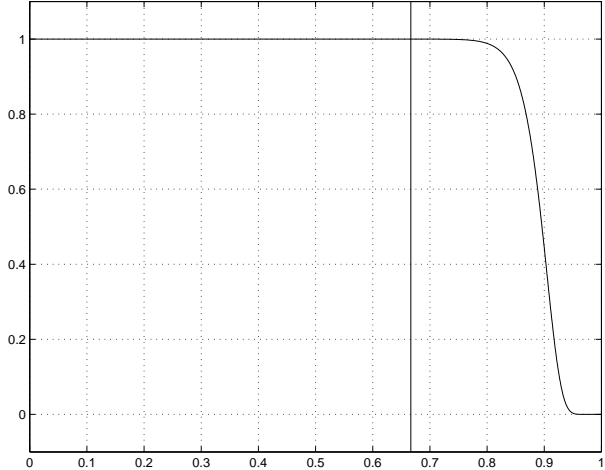}
\end{center}
\caption{The profile of the Fourier smoothing: $\exp(-36 (x)^{36})$.
The vertical line corresponds to the cut-off mode using the $2/3$ 
dealiasing rule. We can see that using this Fourier smoothing we keep 
about $12 \sim 15\%$ more modes than those using the 2/3 dealiasing rule.
\label{fig.fourier_smoother}}
\end{figure}

\begin{figure}
\begin{center}
\includegraphics[width=12cm,height=6cm]{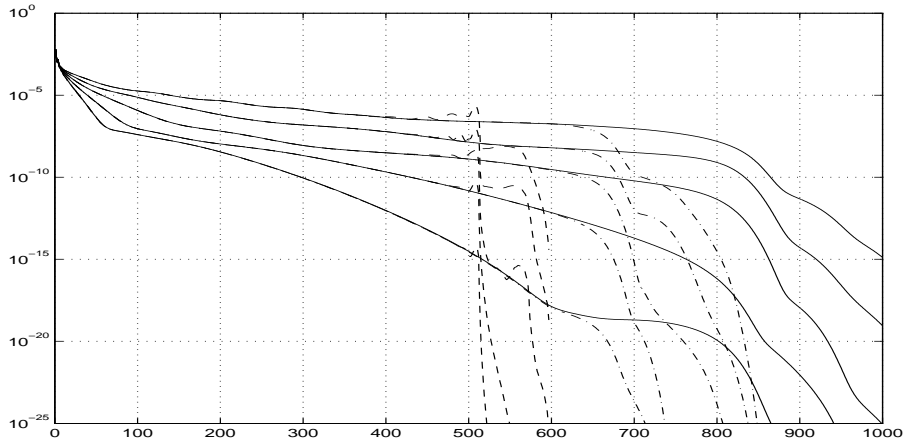}
\end{center}
\caption{The comparison of the enstrophy spectra obtained using the high order 
Fourier smoothing method with those using the 2/3 dealiasing rule. The dashed 
lines and dashed-dotted lines are the enstrophy spectra with the
resolution $768\times 512\times 1536$ using the 2/3 dealiasing rule and the
 Fourier smoothing, respectively. The solid lines are the enstrophy spectra
with resolution $1024\times 768\times 2048$. The times for the spectra lines 
are at $t= 15, 16, 17, 18, 19$ respectively. 
\label{fig.enstrophy-spec-comp}}
\end{figure}

There is a good reason why we choose to use the pseudo-spectral 
method with the above high order Fourier smoothing instead of using 
the classical 2/3 dealiasing rule. The Fourier smoothing we use is 
designed to keep the majority of the Fourier modes unchanged 
and remove the very high modes to avoid the aliasing errors, see 
Fig. \ref{fig.fourier_smoother} for the profile of $\rho (x)$. We 
choose $\alpha$ to be $36$ to guarantee that $\rho (k)$ reaches 
the level of the round-off errors ($O(10^{-16})$) at the highest modes. 
We choose the order of smoothing, $m$, to be 36 in order to optimize 
the accuracy of the spectral approximation, while still keeping the 
aliasing errors under control. As we can see from Figure
\ref{fig.fourier_smoother}, the effective modes in our computation 
are about $12 \sim 15\%$ more than those using the standard $2/3$
dealiasing rule. To demonstrate that the pseudo-spectral method 
with the above high order Fourier smoothing is indeed more accurate 
than the pseudo-spectral method with the $2/3$ dealiasing rule, we 
perform resolution study of the two approaches. In Figure 
\ref{fig.enstrophy-spec-comp}, we compare the Fourier spectra 
of the enstrophy obtained by using the pseudo-spectral method 
with the $2/3$ dealiasing rule with those obtained by the 
pseudo-spectral method with the high order smoothing. For
a fixed resolution $768\times 512\times 1536$, we can see
that the Fourier spectra obtained by the pseudo-spectral
method with the high order smoothing are more accurate than 
those obtained by the spectral method using the $2/3$ 
dealiasing rule. This can be seen by comparing the results
with the corresponding computations using a higher resolution 
$1024 \times 768 \times 2048$. Moreover, the pseudo-spectral
method using the high order Fourier smoothing does not
give the spurious oscillations in the Fourier spectra which 
are present in the computations using the $2/3$ dealiasing 
rule near the $2/3$ cut-off point.

We have used a sequence of resolutions in our computations in
order to perform a resolution study. The resolutions presented in the 
paper are $768\times 512\times 1536$, $1024\times 768\times 2048$,
and $1536\times 1024\times 3072$ respectively. Except for the computation 
on the largest resolution $1536\times 1024\times 3072$, all computations
are carried out from $t=0$ to $t=19$. The computation on the final 
resolution $1536\times 1024\times 3072$ is started from $t=10$ with 
the initial condition given by the computation with the resolution 
$1024\times 768\times 2048$. From our resolution study for the velocity 
and vorticity in both physical and spectral spaces, we find that 
the solution at $t=10$ is fully resolved even on the resolution 
$768\times 512\times 1536$. Thus it is justified to start the computation 
with the largest resolution at $t=10$ using the computation obtained 
by the resolution $1024\times 768\times 2048$ as the initial condition.  

We also exploit the symmetry properties of the solution in our computations,
and perform our computations on only a quarter of the whole domain.
We use the well-known parallel FFT package, called FFTW 3.1 to perform 
the sine and cosine transformations. The whole program is coded in C 
and we use MPI as the parallel interface. Since the 
solution appears to be most singular in the $z$ direction, we allocate twice
as many grid points along the $z$ direction than along the $x$ direction. 
The solution is least singular in the $y$ direction. We allocate the
smallest resolution in the $y$ direction to reduce the computation cost. 
In our computations, two typical ratios in the resolution along the
$x$, $y$ and $z$ directions are $3:2:6$ and $4:3:8$.

In the early stage of the computations for $0 \le t \le 8$, the solution is
still very regular. The solution in this stage can be fully resolved 
using the resolution $768\times 512\times 1536$. In the intermediate stage 
for $8 \le t \le 16$, the solution has a fast growth in the maximum vorticity. 
After $t=16$, the two vortex tubes become severely flattened, and evolve into 
two thin vortex sheets, which roll up subsequently. The maximum vorticity
also experiences a transition from the double exponential growth to
a slower growth rate due to the dynamic depletion of vortex stretching.
In order to resolve the later stage of the solution, higher resolutions
become necessary.

Our computations were carried out on the PC cluster LSSC-II in the 
Institute of Computational Mathematics and Scientific/Engineering Computing 
of Chinese Academy of Sciences and the Shenteng 6800 cluster in the Super 
Computing Center of Chinese Academy of Sciences. The maximal memory consumption 
in our computations is about 120 GBytes. 

\section{Numerical Results}

\subsection{Review of Kerr's results}

In \cite{Kerr93}, Kerr presented numerical evidence which suggested
a finite time singularity of the 3D Euler equations for two perturbed 
antiparallel vortex tubes. He used a 
pseudo-spectral discretization in the $x$ and $y$ directions, and 
a Chebyshev method in the $z$ direction with resolution of order 
$512\times 256 \times 192$. His computations showed that the growth 
of the peak vorticity, the peak axial strain, and the enstrophy 
production obey $(T-t)^{-1}$ with $T = 18.9$. As $t$ approaches 
the alleged blow-up time $T$, the region bounded by the contour 
of $0.6\|\vec{\omega}\|_\infty$, also known as the inner region,  
looks like two vortex sheets with thickness $\sim (T-t)$ 
meeting at an angle \cite{Kerr93}. This region has the length 
scale $(T-t)^{1/2}$ in the vorticity direction. The maximum 
vorticity resides in the small tube-like region, with scaling
$(T-t)^{1/2}\times (T-t)\times (T-t)$, which is the intersection 
of the two sheets. Inside the inner region, the vortex lines 
are ``relatively straight'' \cite{Kerr93}. Kerr stated in his 
paper \cite{Kerr93} (see page 1727) that his numerical results shown 
after $t=17$ and up to $t=18$ were ``not part of the primary evidence 
for a singularity'' due to the lack of sufficient numerical 
resolution and the presence of noise in the numerical solutions. 

In his recent paper \cite{Kerr04} (see also \cite{Kerr97,Kerr99}), 
Kerr applied a high wave number filter to the data obtained in his 
original computations to ``remove the noise that masked the structures 
in earlier graphics'' presented in \cite{Kerr93}. With this filtered 
solution, he presented some scaling analysis of the numerical solutions 
up to $t=17.5$. Two new properties were presented in this recent 
paper \cite{Kerr04}. First, the velocity field was shown to blow up 
like $O(T-t)^{-1/2}$ with $T$ being revised to $T=18.7$. The maximum 
velocity is located on the boundary of the inner region with 
distance $(T-t)^{1/2}$ away from the position where the maximum 
vorticity is achieved. Secondly, he showed that the blowup is 
characterized by two anisotropic length scales, 
$\rho \approx (T-t) $ and $R \approx (T-t)^{1/2}$. 
Further, Kerr argued that the curvature, $\kappa$, of the vortex
lines and $\nabla\cdot \vec{\xi}$ (here
$\vec{\xi} =\vec{\omega}/|\vec{\omega}|$) in 
the inner region are likely bounded 
by $(T-t)^{-1/2}$ \cite{Kerr04}. 

\subsection{Recent theoretical results on the 3D Euler equations}

There has been some interesting development in the theoretical 
understanding of the 3D incompressible Euler equations. It has been 
shown by several authors that the local geometric regularity of vortex 
lines can play an important role in depleting nonlinear vortex stretching. 
In particular, Constantin, Fefferman and Majda \cite{CFM96} proved that if 
(i) there is up to time $T$ an $O(1)$ region $\Omega$ in which the vorticity 
vector is smoothly directed, i.e., 
the maximum norm of $\nabla \xi$ in this $O(1)$ region 
is $L^2$ integrable in time,
(ii) the maximum norm of velocity is uniformly bounded in time, plus 
a technical condition on the distribution of the vorticity within this 
region, then no blow-up can occur in this region up to time $T$. While 
this result is very interesting, it does not apply to Kerr's computations 
since the two assumptions are violated by Kerr's computations.

Motivated by the result of \cite{CFM96}, Deng, Hou, and Yu \cite{DHY05a} 
obtained sharper non-blowup conditions which use only very localized information
of the vortex lines. Assume that at each time $t$ there exists some vortex line
segment $L_t$ on which the local maximum vorticity is comparable to the global maximum
vorticity. Further, we denote $L(t)$ as the arclength of $L_t$.
Roughly speaking, if (i) the velocity field along $L_t$ is bounded
by $C_U (T-t)^{-\alpha}$ for some $\alpha < 1$, (ii) 
$C_L (T-t)^\beta \le L(t) \le C_0 /\max_{L_t}(|\kappa|,\;|\nabla \cdot \vec{\xi}|)$, 
for some $\beta < 1 - \alpha$, then the 
solution of the 3D Euler equations remains regular up to $T$. In Kerr's
computations, the first condition is satisfied with $\alpha = 1/2$. Moreover,
based on the bound on $\kappa$ and $\nabla \cdot \vec{\xi}$ in the inner 
region \cite{Kerr04}, we can choose a vortex line segment of length 
$(T-t)^{1/2}$ in the inner region so that the upper bound in the
second condition is satisfied. However, the lower bound is violated 
since $\beta < 1/2 = 1 - \alpha $. In a subsequent paper \cite{DHY05b}, 
the non-blowup conditions have been improved to include the critical case, 
$\beta = 1 - \alpha$ by requiring the scaling constants, $C_U, \;C_0,$ and 
$C_L$ in conditions (i)-(ii), to satisfy an algebraic inequality. 
This algebraic inequality can be checked numerically if we obtain a good 
estimate of these scaling constants. For example, if $C_0 = 0.1$, which 
seems reasonable since the vortex lines are relatively straight in the 
inner region, the result of \cite{DHY05b} implies no blowup up to 
$T$ if $2C_U < 0.43 C_L$. However, such scaling constants are not 
available in \cite{Kerr93}. One of our original motivations was 
to repeat Kerr's computations with higher resolution to obtain a good 
estimate of these scaling constants.

\subsection{Maximum vorticity growth}

\begin{figure}
\begin{center}
\includegraphics[width=8cm]{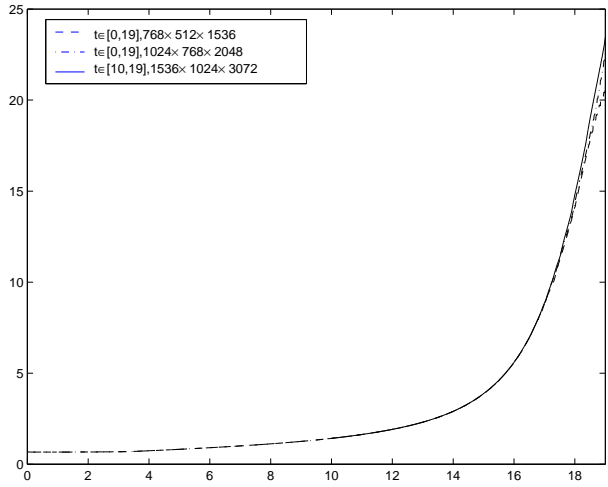}
\end{center}
\caption{The maximum vorticity $\|\omega\|_\infty$ in time using 
different resolutions. 
\label{fig.omega}}
\end{figure}

We first present the result on the growth of the maximum vorticity in time,
see Figure \ref{fig.omega}. The maximum vorticity increases rapidly from 
the initial value of $0.669$ to $23.46$ at the final time $t=19$, a factor 
of 35 increase from its initial value. Kerr's computations predicted a 
finite time singularity at $T=18.7$. Our computations show no sign of finite 
time blowup of the 3D Euler equations up to $T=19$, beyond the singularity 
time predicted by Kerr. In Figure \ref{fig.omega},
we plot the maximum vorticity in time using three different 
resolutions, i.e. $768\times 512\times 1536$, $1024 \times 768 \times 2048$,
and $1536 \times 1024 \times 3072$ respectively. As we can see, the
agreement between the two successive resolutions is very good with 
only mild disagreement toward the end of the computations. This indicates 
that a very high space resolution is indeed needed to capture the rapid 
growth of maximum vorticity at the later stage of the computations. 

\begin{figure}
\begin{center}
\includegraphics[width=8cm]{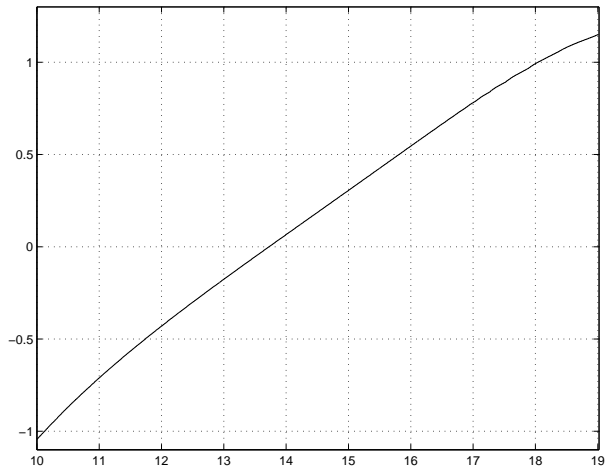}
\end{center}
\caption{The plot of $ \log \log \|\omega\|_\infty$ vs time, resolution $1536\times 1024
\times 3072$. 
\label{fig.omega_loglog}}
\end{figure}

\begin{figure}
\begin{center}
\includegraphics[width=8cm]{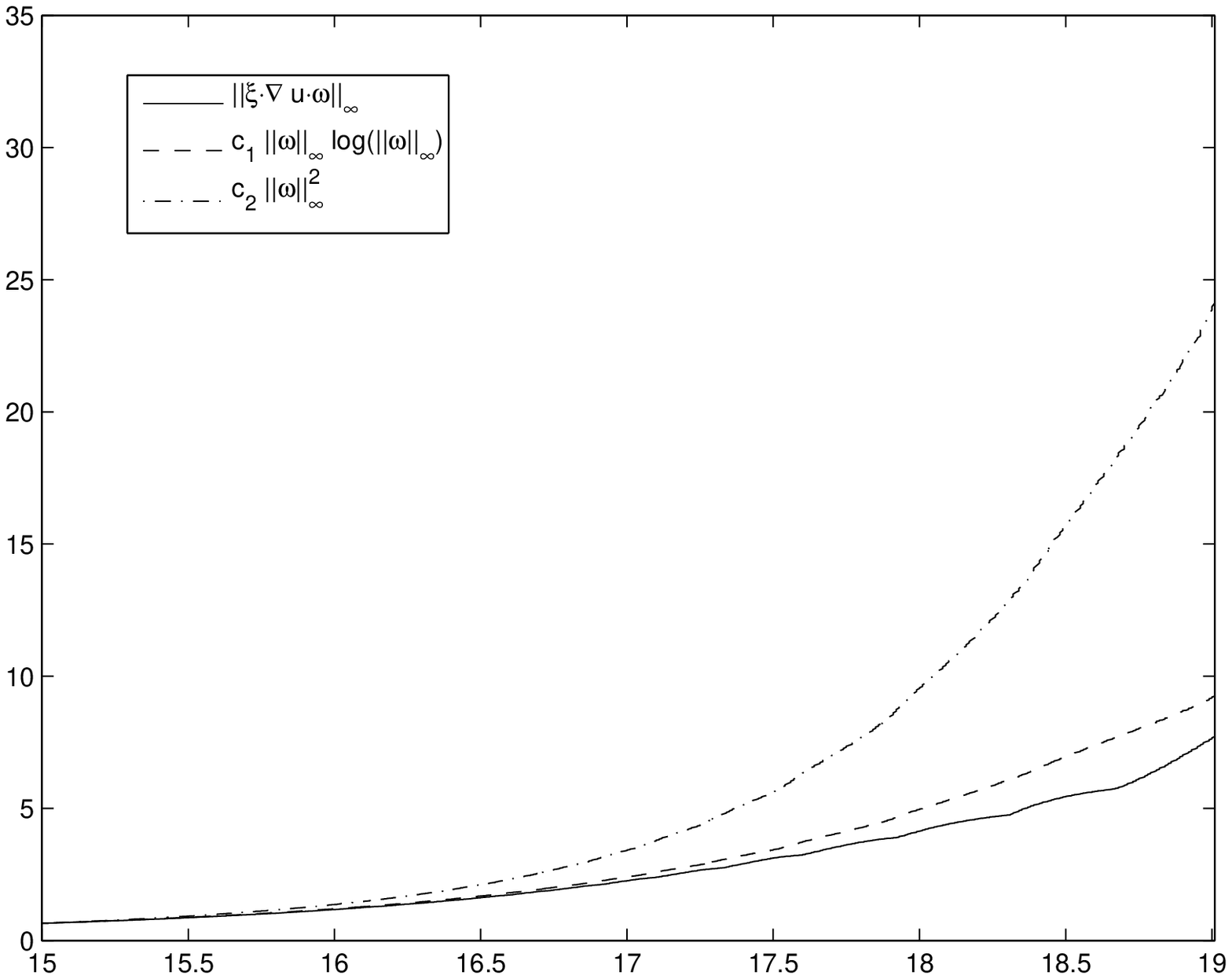}
\end{center}
\caption{Study of the vortex stretching term in time, resolution $1536\times 
1024\times 3072$. We take $c_1 = 1/8.128$, $c_2 = 1/23.24$ to match the 
same starting value for all three plots. 
\label{fig.growth_rate}}
\end{figure}

We observe that the growth of the maximum vorticity has three distinguished 
phases. The first stage is for $0\le t \le 12$. In this early stage, 
the maximum vorticity 
grows only exponentially in time. This is consistent with Kerr's results. 
The second stage is for $12 \le t\le 17$. During this intermediate stage,
the two vortex tubes experience tremendous core deformation and become 
severely flattened. Each vortex tube effectively turns into a vortex sheet 
with rapidly decreasing thickness. We observe that the growth of
maximum vorticity is slightly slower than double exponential in time 
during the second stage, see Figure \ref{fig.omega_loglog}. This growth 
behavior can be also confirmed by examining the degree of nonlinearity 
in the vortex stretching term. If the maximum vorticity indeed blew up 
like $O(T-t)^{-1}$, as alleged in \cite{Kerr93}, the vortex stretching term 
at the position of the maximum vorticity should have been quadratic as a 
function of maximum vorticity. However, as Figure \ref{fig.growth_rate} 
shows, the vortex stretching term, when projected to the unit vorticity 
vector, grows much slower than the quadratic nonlinearity. In fact, it 
is even slower than 
$C\|\vec{\omega} \|_\infty \log (\|\vec{\omega} \|_\infty )$, i.e.
\begin{equation}
\label{eqn:stret}
\| \vec{\xi} \cdot \nabla \vec{u} \cdot \vec{\omega} \|_\infty
\le C \|\vec{\omega} \|_\infty \log (\|\vec{\omega} \|_\infty ), \quad \quad
15 \le t \le 19 .
\end{equation}
Using the equation that governs the magnitude of vorticity \cite{Const94}, 
\begin{equation}
\frac{\partial}{\partial t} |\vec{\omega}| + (\vec{u}\cdot\nabla) |\vec{\omega}| 
 = \vec{\xi} \cdot \nabla \vec{u} \cdot \vec{\omega} ,
\end{equation}
one can easily show that inequality (\ref{eqn:stret}) implies that
the maximum vorticity is bounded by double exponential in time.

During the final stage for $17 \le t \le 19$, we observe that the growth of
the maximum vorticity slows down and deviates from double exponential growth, 
see Figure \ref{fig.omega_loglog}. This indicates that there is stronger 
cancellation taking place in the vortex stretching term.

\begin{figure}
\begin{center}
\includegraphics[width=8cm]{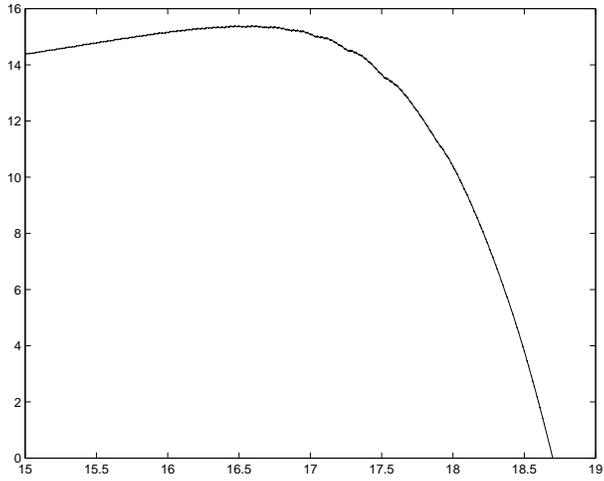}
\end{center}
\caption{Scaling constant in time for the fitting $\|\omega\|_\infty \approx c/(T-t)$,
$T=18.7$.\label{fig.scaling_const}}
\end{figure}

We remark that for vorticity that grows as rapidly as double exponential in time,
one may be tempted to fit the maximum vorticity growth by $(T-t)^{-1}$ for some 
$T$. Indeed, if we choose $T=18.7$ as suggested by Kerr in \cite{Kerr04}, we
find a reasonably good fit for the maximum vorticity as a function of
$c/(T-t)$ for the period $15 \le t \le 17$. We plot the scaling constant
$c$ in Figure \ref{fig.scaling_const}. As we can see, $c$ is close to a 
constant for $15 \le t \le 17$. To conclude that the 3D Euler equations
indeed develop a finite time singularity, one must demonstrate that such 
scaling persists as $t$ approaches to $T$. As we can see from Figure 
\ref{fig.scaling_const}, the scaling constant $c$ decreases rapidly to
zero as $t$ approaches to the alleged singularity time $T$. Therefore,
the fitting of $\| \vec{\omega}\|_\infty \approx O(T-t)^{-1}$ is not correct
asymptotically.

\begin{figure}
\begin{center}
\includegraphics[width=8cm]{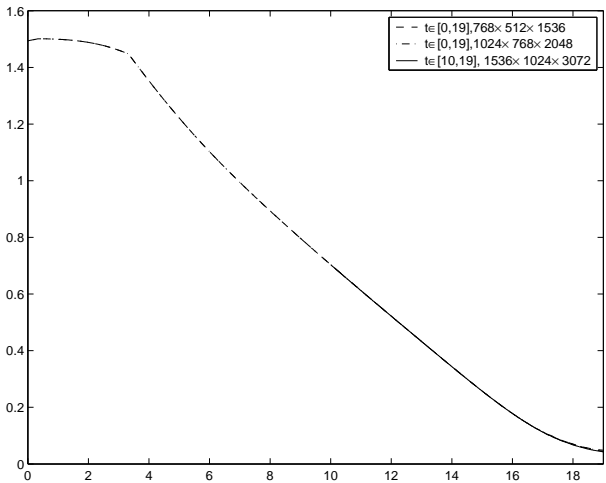}
\end{center}
\caption{The inverse of maximum vorticity $\|\omega\|_\infty$ in time 
using different resolutions. 
\label{fig.omega1}}
\end{figure}

A similar test can be performed for the inverse of the maximum vorticity.
In Figure \ref{fig.omega1}, we plot the inverse of the maximum vorticity
using different resolutions. As we can see from this picture, the 
inverse of the maximum vorticity approaches to zero almost linearly 
in time for $8 \le t \le 17$. This was one of the strong evidences 
presented in \cite{Kerr93} that suggests a finite time blowup of the 
3D Euler equations. If this trend were to continue to hold up to $T$, 
it would have led to the blowup of the maximum vorticity in the form of 
$O(T-t)^{-1}$. However, as we increase our resolutions, we find that the 
curve corresponding to the inverse of the maximum vorticity starts to turn 
away from zero around $t=17$. This is precisely the time when Kerr's 
computations began to lose resolution. By $t=17.5$, the gradients of the
solution become very large in all three directions. In order to resolve the 
nearly singular solution structure, we use $1536\times 1024\times 3072$ grid 
points from $t=10$ to $19$. This level of resolution gives about 8 grid points 
across the most singular region in each direction toward the end of the
computations.

\subsection{Velocity profile}

\begin{figure}
\begin{center}
\includegraphics[width=8cm]{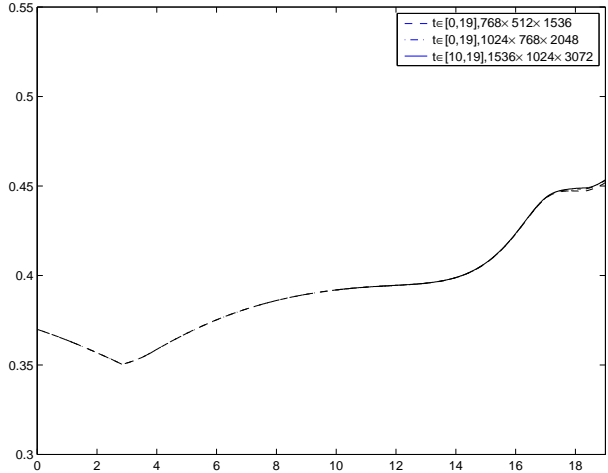}
\end{center}
\caption{Maximum velocity $\|\vec{u}\|_\infty$ in time using
different resolutions.\label{fig.velocity}}
\end{figure}

\begin{figure}
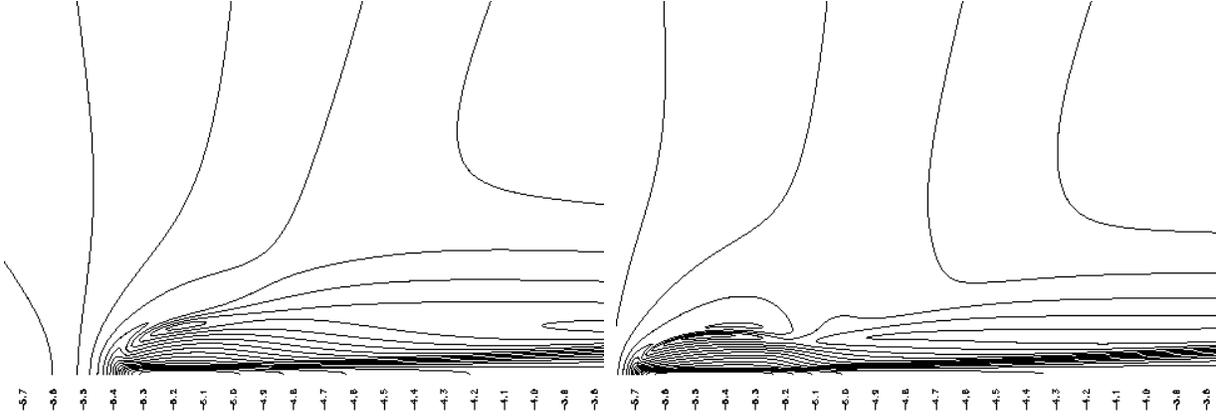

\begin{center}
\includegraphics[width=8cm]{pics/velo_cont-t=18.epsf}
\includegraphics[width=8cm]{pics/velo_cont-t=19.epsf}
\end{center}
\caption{The contour of $|\vec{u}|$ in the region of maximum vorticity
on the symmetry plane. The pictures are plotted at $t=18$ and $t=19$ 
respectively using resolution $1536\times 1024\times 3072$. 
\label{velo-cont}}
\end{figure}

One of the important findings of our computations is that the velocity 
field is actually bounded by 1/2 up to $T=19$. This is in contrast to Kerr's 
computations in which the maximum velocity was shown to blow up like 
$O(T-t)^{-1/2}$. We plot 
the maximum velocity as a function of time using different 
resolutions in Figure \ref{fig.velocity}. The computation 
with the resolution $768\times 512\times 1536$ shows some 
mild discrepancy toward the end of the computation. On the
other hand, the computation obtained by resolution
$1024\times 768 \times 2048$ and the one obtained
by resolution $1536\times 1024\times 3072$ are
almost indistinguishable. As we can see, the maximum velocity 
grows slowly in time and is relatively
small in magnitude. There is a relatively fast growth of maximum velocity 
between $t=14$ and $t=17$. But this growth becomes saturated by $t=17.5$. 
After $t=18.4$, the velocity experiences a mild growth, but it is still 
bounded by $0.46$ at the final time $T=19$. 
We also plot the contours
of $|\vec{u}|$ near the region of maximum vorticity 
at $t=18$ and $19$ in Figure \ref{velo-cont}. As we can see,
the velocity seems to be well resolved.

%

The fact that the velocity field is bounded is significant. 
By re-examining the non-blowup conditions of the theory of 
Deng-Hou-Yu \cite{DHY05a}, we find that the first condition is 
now satisfied with $\alpha = 0$ since the velocity field is bounded.
According to \cite{Kerr04}, we have
$\max_{L_t}(|\kappa|,\;|\nabla \cdot \vec{\xi}|) \le C_0 (T-t)^{-1/2}$.
In fact, our computations indicate that the curvature and the 
divergence of the unit vorticity vector are actually bounded. With 
$\alpha = 0$, we can now choose the vortex line segment of
length, $L(t) = (T-t)^\beta $ with $\beta =1/2 < 1-\alpha$, so that 
the second condition is now satisfied. Thus, the theory of 
Deng-Hou-Yu applies, which implies non-blowup of the 3D Euler 
equations up to $T$.

\subsection{Local vorticity structure}

\begin{figure}
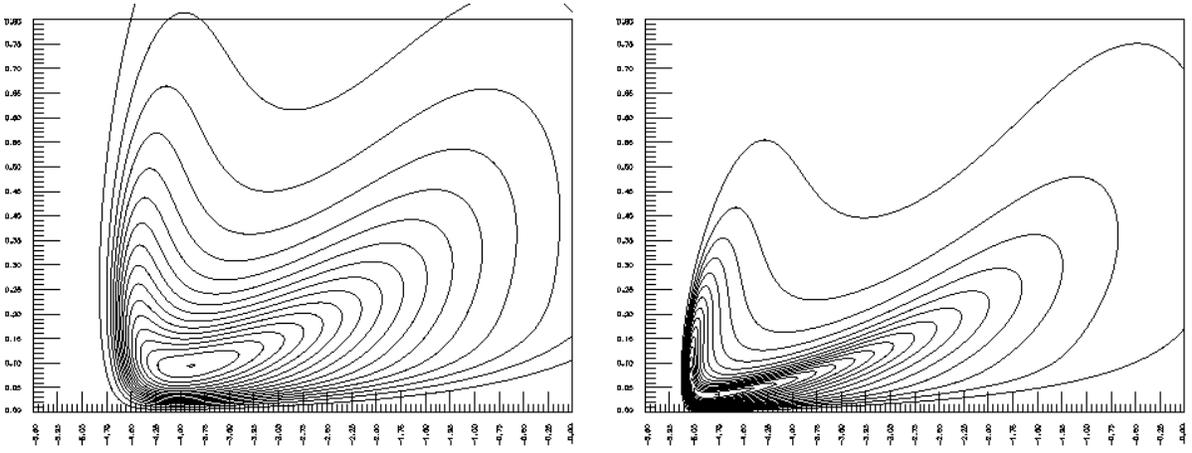

\begin{center}
\includegraphics[width=8cm]{pics/scaled_contour_t=15.epsf}
\includegraphics[width=8cm]{pics/scaled_contour_t=17.epsf}
\end{center}
\caption{The contour of axial vorticity around the maximum vorticity on the symmetry 
plane at $t=15, 17$. The figure is scaled in $z$ direction by a factor of $5$ to 
compare with the Figure 4 in \cite{Kerr93}. A contour at value very close to the maximum
value is plotted to show the location of the maximum vorticity. \label{fig.scaled_local_struc}}
\end{figure}

\begin{figure}
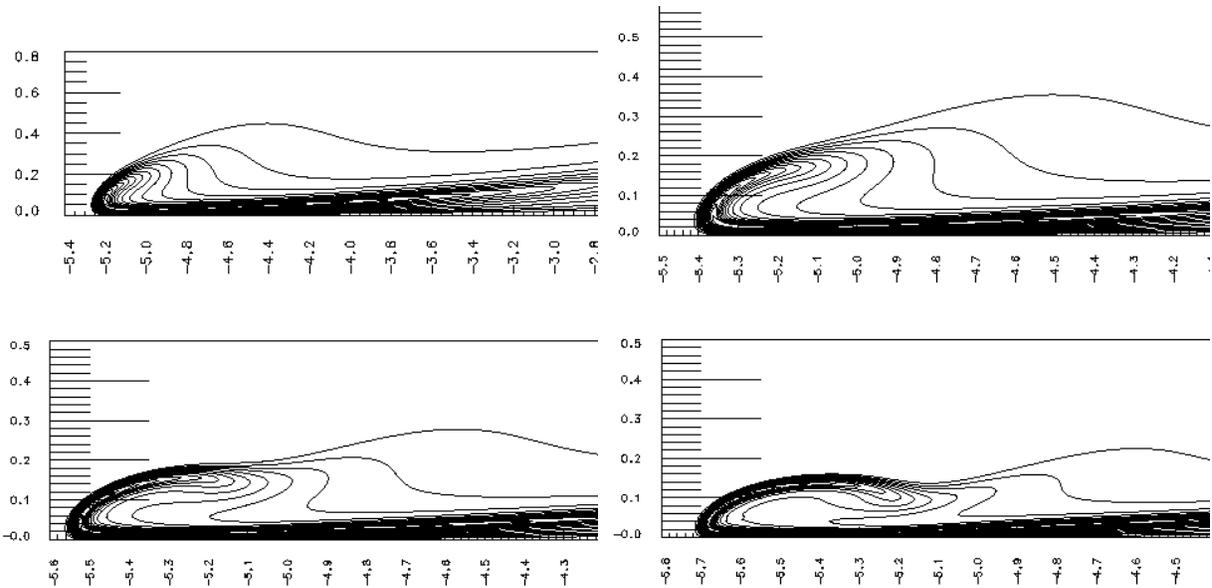

\begin{center}
\includegraphics[width=8cm]{pics/fig4.noscale.t=17.5.epsf}
\includegraphics[width=8cm]{pics/fig4.noscale.t=18.epsf}
\includegraphics[width=8cm]{pics/fig4.noscale.t=18.5.epsf}
\includegraphics[width=8cm]{pics/fig4.noscale.t=19.epsf}
\end{center}
\caption{The contour of axial vorticity around the maximum vorticity on
the symmetry plane at $t=17.5, 18, 18.5, 19$. \label{fig.local_struc}}
\end{figure}

In this subsection, we would like to examine the local vorticity structure
near the region of the maximum vorticity. To illustrate the development in
the symmetry plane, we show a series of vorticity contours near the region
of the maximum vorticity at late times in a manner similar to the results
presented in
\cite{Kerr93}. In Kerr's computations, he observed that the head and tail 
in the symmetry plane develop a corner separating the head and tail. We 
adopt Kerr's definition of the ``head'' to be the region extending above 
the vorticity peak $\omega_p$ just behind the leading edge of the vortex
sheet. 
The ``tail'' is the vortex sheet extending behind the peak vorticity. 
One interesting question is to determine whether one direction becomes 
progressively more flattened or stretched as the flow evolves and whether 
the rates of collapse are the same in different directions. Our 
computational results are in qualitative agreement with Kerr's in 
the early and intermediate stages. In particular, we observe that as the
flow evolves the region of peak vorticity concentrates into the region
where the vortex sheets of the head and tail meet. To compare with Kerr's 
figures, we scale the vorticity contours in the $x-z$ plane by a 
factor of 5 in the $z$ direction. The results at $t=15$ and $t=17$ are 
plotted in Figure \ref{fig.scaled_local_struc}. We can see that the
location of maximum axial vorticity moves toward the corner where
the vortex sheets of the head and tail meet as time increases,
see also Figure \ref{fig.local_struc}.
This is in qualitative agreement with Kerr's results.

In order to see better the dynamic development of the local vortex structure,
we plot a sequence of vorticity contours on the symmetry plane at 
$t=16, 17, 18,$ and $19$ respectively in Figure \ref{fig.local_struc}. 
The pictures are plotted using the
original length scales, without the scaling by a factor of 5 in the $z$ 
direction as in Figure \ref{fig.scaled_local_struc}. From these results, we 
can see that the vortex sheet is compressed in the $z$ direction. 
It is clear that a thin layer (or a vortex sheet) is formed dynamically. 
The head of the vortex sheet is a bit thicker 
than the tail at the beginning. The head of the vortex 
sheet begins to roll up around $t=16$. By the time $t=19$, the 
head of the vortex sheet has traveled backward for quite a distance, 
and the vortex sheet has been compressed quite strongly along 
the $z$ direction.

\begin{figure}
\begin{center}
\includegraphics[width=10cm]{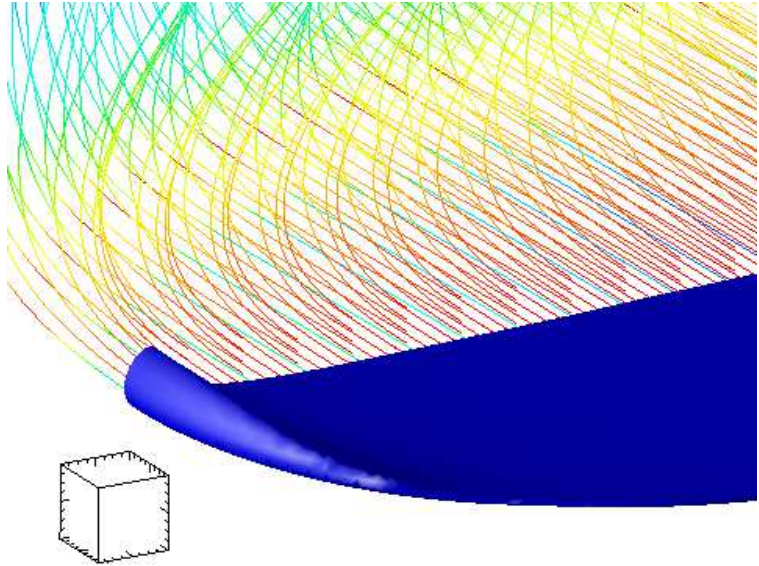}
\end{center}
\caption{The local 3D vortex structure and vortex lines around the maximum
vorticity at $t=17$. The size of the box on the left is 
$0.075^3$ to demonstrate the scale of the picture. 
\label{fig.local_struc_3d_17}}
\end{figure}

\begin{figure}
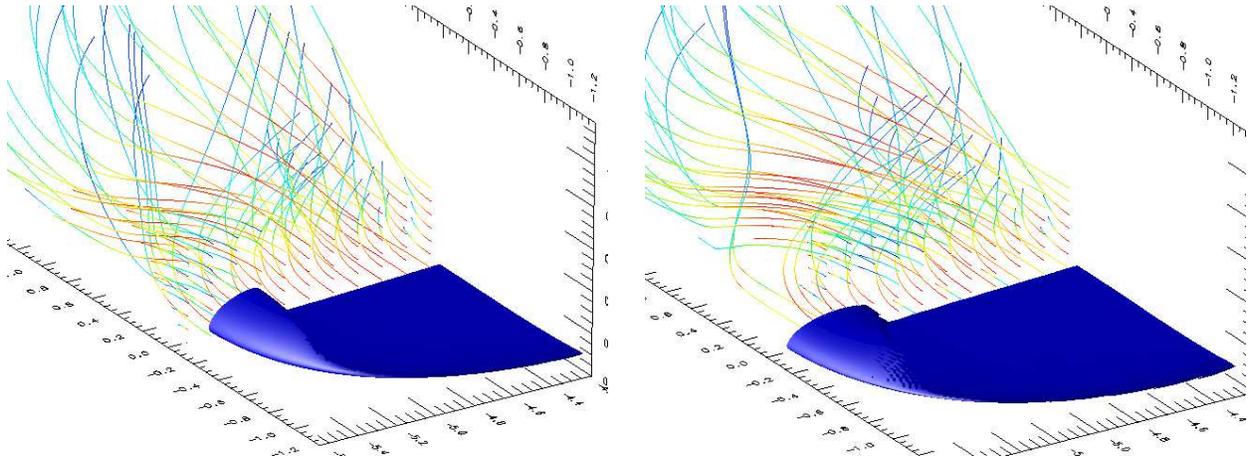

\begin{center}
\includegraphics[width=8cm]{pics/l-18.epsf}
\hspace{2mm}
\includegraphics[width=8cm]{pics/l-19.epsf}
\end{center}
\caption{
The local 3D vortex structures and vortex lines around the maximum
vorticity at $t=18$ (on the left) and $t=19$ (on the right). 
\label{fig.local_struc_3d}}
\end{figure}

We also plot the isosurface of vorticity near the region of the maximum 
vorticity in Figures \ref{fig.local_struc_3d_17} and \ref{fig.local_struc_3d} 
to illustrate the dynamic roll-up of the vortex sheet near the region of the 
maximum vorticity. Figure \ref{fig.local_struc_3d_17} gives 
the local vorticity structure at $t=17$. If we scale the local 
roll-up region on the left hand side next to the box by a factor of 4 
along the $z$ direction, as was done in \cite{Kerr04}, we would obtain a 
local roll-up structure which is qualitatively similar to 
Figure 1 in \cite{Kerr04}. 
In Figure \ref{fig.local_struc_3d}, we show the local vorticity structure
for $t=18$ and $19$. In both figures, the isosurface
is set at $0.5 \times \|\vec{\omega}\|_\infty$. Here we make a 
few observations. First, the vortex lines near the region of 
maximum vorticity are relatively straight and the vorticity vectors 
seem to be quite regular. This was also observed in \cite{Kerr93}. 
On the other hand, the inner region containing the maximum vorticity 
does not seem to 
shrink to zero at a rate of $(T-t)^{1/2}$, as predicted in \cite{Kerr93}.
The length and the width of the vortex sheet are still 
$O(1)$, although the thickness of the vortex sheet becomes quite small. 

\begin{figure}
\begin{center}
\includegraphics[width=8cm]{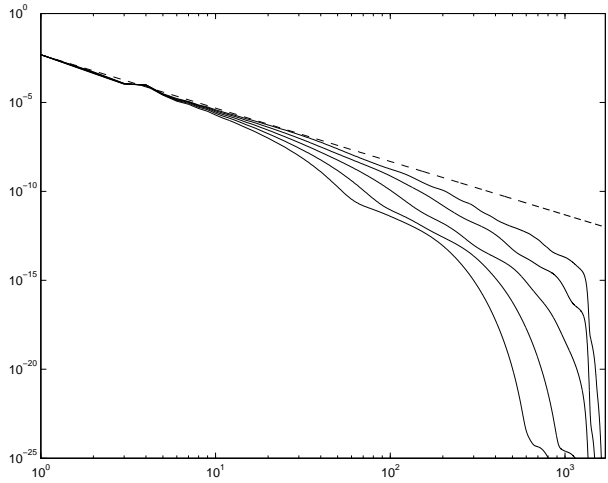}
\end{center}
\caption{The energy spectra for velocity at different times in log-log scale. 
The energy spectrum is calculated using $E(k) = \sum_{|\vec{k}|\in (k-1/2, k+1/2]} 
|\widehat{u}_{\vec{k}}|^2$. The time for the spectral lines from bottom to top are $t=
15, 16, 17, 18, 19$. The dashed line corresponds to $k^{-3}$. 
\label{fig.energy_spec}}
\end{figure}

\begin{figure}
\begin{center}
\includegraphics[width=8cm]{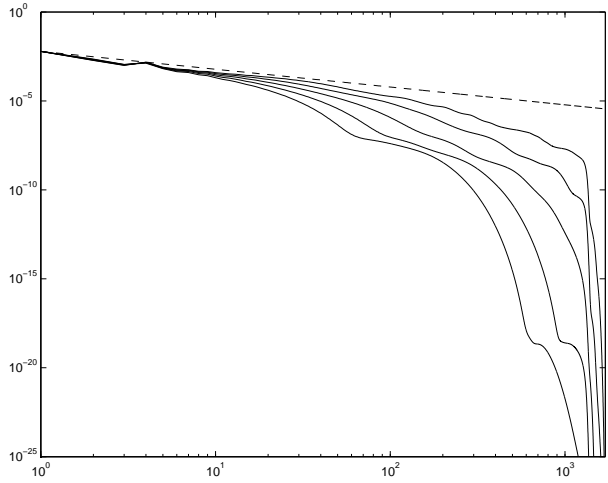}
\end{center}
\caption{The enstrophy spectra for vorticity at different times in log-log scale.
The enstrophy spectrum is calculated using $Ens(k) = \sum_{|\vec{k}|\in (k-1/2, k+1/2]}
|\widehat{\omega}_{\vec{k}}|^2$. The dashed line corresponds to $k^{-1}$. 
\label{fig.enstrophy_spec}}
\end{figure}

\begin{figure}
\begin{center}
\includegraphics[width=8cm]{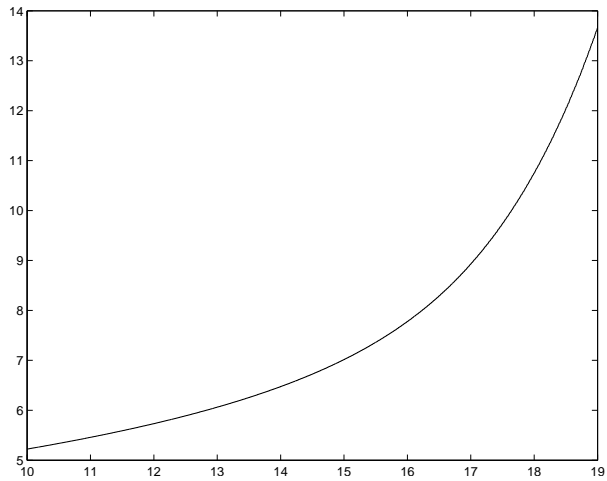}
\end{center}
\caption{The enstrophy as a function of time, resolution $1536\times 1024\times 3072$.
\label{fig.omegal2}}
\end{figure}

\begin{figure}
\begin{center}
\includegraphics[width=8cm]{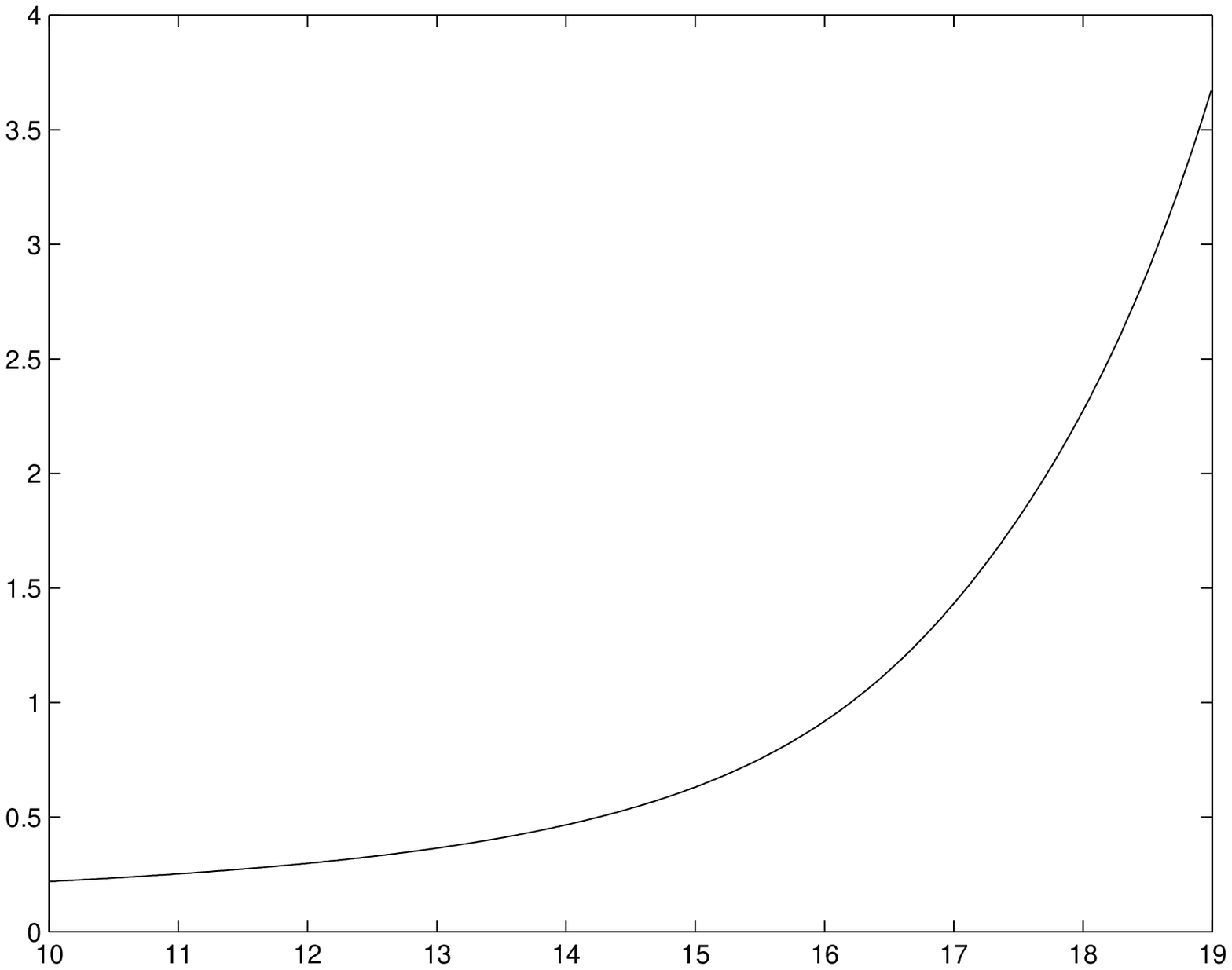}
\includegraphics[width=8cm]{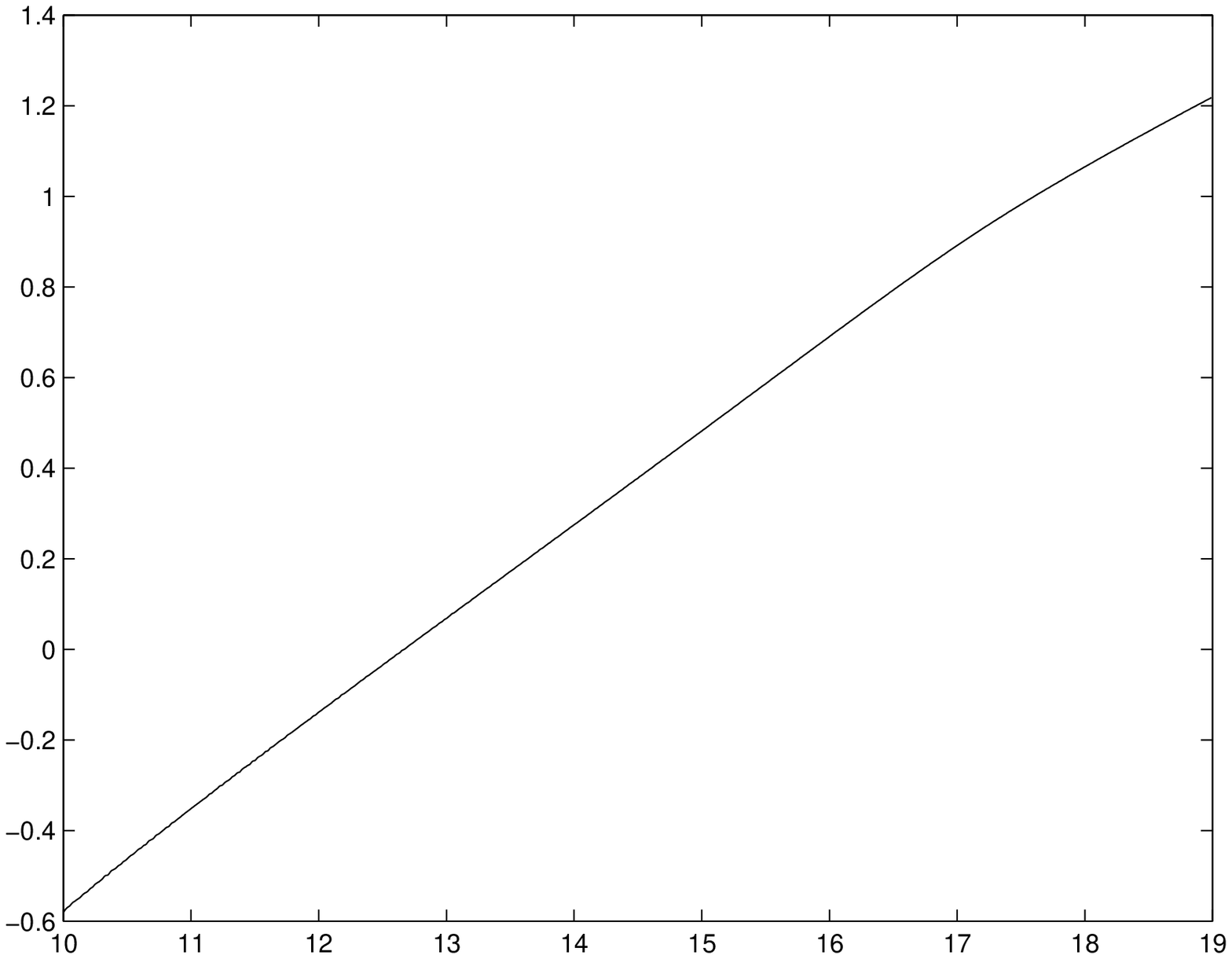}
\end{center}
\caption{The enstrophy production rate (left) and the double logarithm of 8 
times enstrophy production rate (right) as a function of time, resolution 
$1536\times 1024\times 3072$. 
\label{fig.enstrophy_production}}
\end{figure}

We also plot the energy spectrum in Figure \ref{fig.energy_spec} at
$t=16, 17, 18, 19$. A finite 
time blow-up of enstrophy would imply that the energy spectrum decays no
faster than $|k|^{-3}$. In \cite{Kerr93}, Kerr observed that the
energy spectrum decays exactly like $|k|^{-3}$, suggesting a finite time
blow-up of the enstrophy (recall that enstrophy is defined as 
the square of the $L^2$ norm of vorticity, i.e.
$\|\vec{\omega}\|_2^2$). 
Our computations show that the energy spectrum
approaches $|k|^{-3}$ for $|k| \le 100$ as time increases to $t=19$.
This is in qualitative agreement with Kerr's results. Note that
there are only less than 100 modes available along the $|k_x|$ or
$|k_y|$ direction
in Kerr's computations, see Figure 18 (a)-(b) of \cite{Kerr93}.
On the other hand, our computations show that 
the high frequency Fourier spectrum for $100 \le |k| \le 1300$ 
decays much faster than 
$|k|^{-3}$, as one can see from Figures \ref{fig.energy_spec} 
and \ref{fig.energy_spec_1}.
This indicates that there is no blow-up in enstrophy. This is also
supported by the enstrophy spectrum, given in Figure 
\ref{fig.enstrophy_spec}, and the plot of enstrophy 
as a function of time in Figures \ref{fig.omegal2}.
In Figure \ref{fig.enstrophy_production}, we plot the enstrophy 
production rate, which is defined as the time derivative of enstrophy. 
Although it grows relatively fast, it actually grows
slower than double exponential in time (see the picture on 
the right in Figure \ref{fig.enstrophy_production}). In the double 
logarithm plot of the enstrophy production rate, we multiply the 
enstrophy production rate by a constant factor 8 to make the second 
logarithm well-defined. It is interesting to note that 
the double logarithm of enstrophy production rate in Figure 
\ref{fig.enstrophy_production} is qualitatively similar to the double 
logarithm of $||\vec{\omega}||_\infty$ in Figure \ref{fig.omega_loglog}. 

\subsection{Resolution Study}

In this subsection, we perform a resolution study to make sure
that the nearly singular behavior of the 3D Euler equations is 
resolved by our computational grid. In Figure \ref{fig.omega},
we have performed resolution study for the maximum vorticity 
using three different resolutions and found very good agreement 
for the time interval $[0,18]$.
There is only a mild disagreement toward the end of the 
computations from $t=18$ to 19.
In our computations with different resolutions, we find that the 
maximum vorticity always grows slower than double exponential in time.
Similar resolution study has been performed for the inverse of
maximum vorticity in Figure \ref{fig.omega1} and for the
maximum velocity in Figure \ref{fig.velocity}. We observe excellent 
agreement between the solutions obtained by the two largest resolutions. 
We have also performed similar resolution study in the Fourier
space by examining the convergence of the energy and enstrophy 
spectra using different resolutions. We observe that the
Fourier spectrum corresponding to the effective modes of one
resolution is in excellent agreement with that corresponding 
a higher resolution computation, 
see Figures \ref{fig.enstrophy-spec-comp}, 
\ref{fig.enstrophy_spec_1} and \ref{fig.energy_spec_1}.  

\begin{figure}
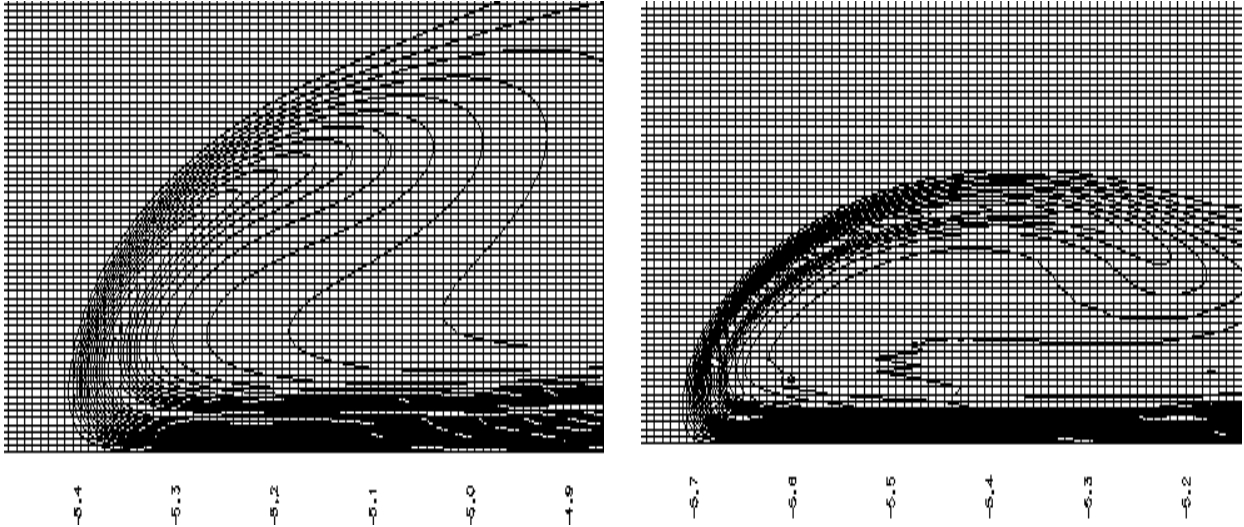

\begin{center}
\includegraphics[width=8cm,height=7cm]{pics/vort_cont_mesh-t=18-1.epsf}
\hspace{2mm}
\includegraphics[width=8cm,height=7cm]{pics/vort_cont_mesh-t=19-1.epsf}
\end{center}
\caption{This picture is to illustrate the mesh around the maximum vorticity. 
The times for this plot are $18, 19$. At $t=19$, we still have about $8$ 
points along $z$ direction to resolve the nearly singular layer.
\label{fig.local_mesh}}
\end{figure}

\begin{figure}
\begin{center}
\includegraphics[width=12cm,height=6cm]{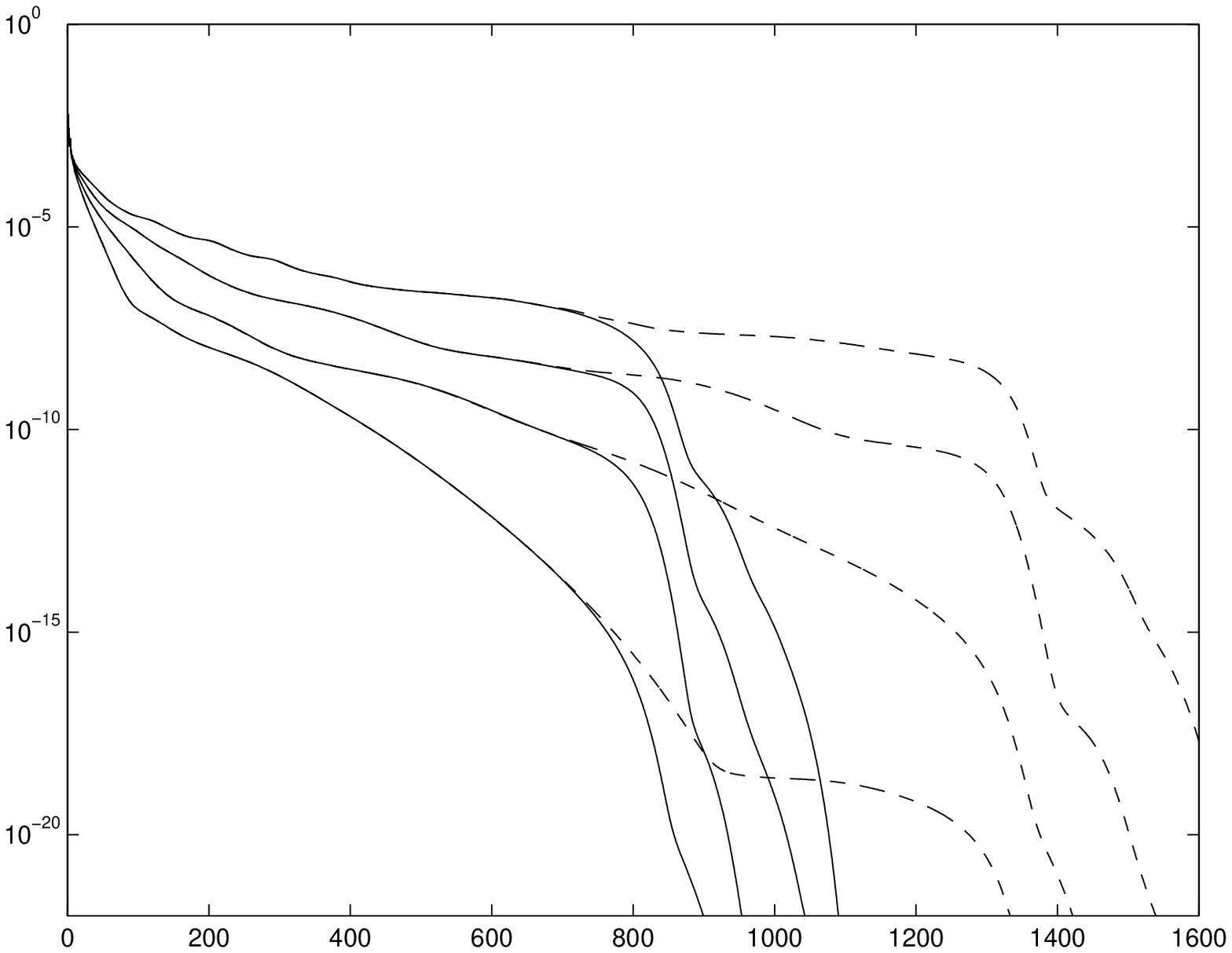}
\end{center}
\caption{Convergence study for enstrophy spectra using different resolutions.
The dashed lines and the solid lines are the enstrophy 
spectra on resolution $1536\times 1024\times 3072$ and $1024\times 768\times
2048$, respectively. The times for the lines from bottom to top are $t=16, 17,
18, 19$.
\label{fig.enstrophy_spec_1}}
\end{figure}

\begin{figure}
\begin{center}
\includegraphics[width=12cm,height=6cm]{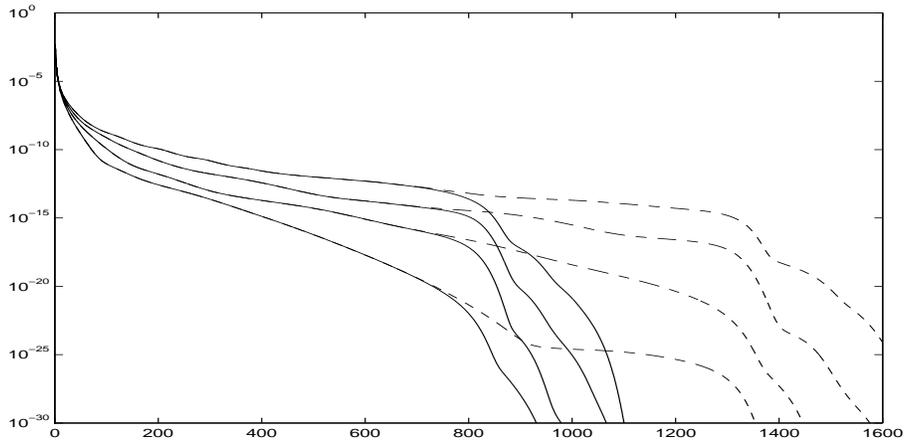}
\end{center}
\caption{Convergence study for energy spectra using different resolutions.
The dashed lines and the solid lines are the energy spectra on resolution
$1536\times 1024\times 3072$ and $1024\times 768\times 2048$, 
respectively. The times for the lines from bottom to top are $t=16, 17,
18, 19$.
\label{fig.energy_spec_1}}
\end{figure}

To see how many grid points we have across the most singular
region, we plot the underlying mesh for the vorticity contours
in the $x-z$ plane in Figure \ref{fig.local_mesh}. One can see 
from this picture that we have about $16$ grid points in the $z$ 
direction at $t=18$ and $8$ grid points at $t=19$. It is also 
interesting to note that at $t=18.5$, the location of the maximum 
vorticity has moved away from the bottom of the vortex sheet 
structure. 
If the current trend continues, it is likely that the location of the 
maximum vorticity will continue to move away from the bottom of the 
vortex sheet. One of the possible blow-up scenarios is that the
interaction of the two perturbed antiparallel vortex tubes would induce
a strong compression between the two vortex tubes, leading to a finite time
collapse of the two vortex tubes. The fact that the location of the
maximum vorticity moves away from the dividing plane of the two vortex 
tubes seems to destroy the desired mechanism to produce a blow-up. 

\section{Concluding Remarks}

We investigate the interaction of two perturbed vortex tubes for the 3D 
Euler equations using Kerr's initial data. Our numerical computations 
demonstrate a very subtle dynamic depletion of vortex stretching. The
maximum vorticity is shown to grow no faster than double exponential
in time up to $T=19$, beyond the singularity time predicted by Kerr in
\cite{Kerr93}. The local geometric regularity of vortex lines seems 
to be responsible for this dynamic depletion of vortex stretching. 
Sufficient numerical resolution is essential in capturing the double 
exponential growth in vorticity and the dynamic depletion
of vortex stretching. The velocity field and the enstrophy are
shown to be bounded throughout the computations. We provide 
evidence that the vortex stretching term is only weakly nonlinear and 
is bounded
by $ \|\vec{\omega} \|_\infty \log (\|\vec{\omega} \|_\infty )$. Such an 
upper 
bound on the vortex stretching term implies that the maximum vorticity 
is bounded by the double exponential in time. Our computational results 
also satisfy the non-blowup conditions of Deng-Hou-Yu, which 
provides a theoretical support for our computational results.

The current computations, even with this level of resolution, 
can not rule out the possibility of the blow-up of the 3D Euler 
equations for large times for Kerr's initial data. The theoretical 
results of \cite{CFM96,DHY05a,DHY05b} and the computations presented 
here suggest that a finite time singularity, if it exists, would have rather
complicated geometric structures. There are other types of potential
Euler singularities that are not considered in this paper. Among them,
the Kida-Pelz initial condition \cite{BP94,Pelz98} is worth further
investigation.
The extra symmetry constraints in this type of initial data are believed
to be important in producing a finite time singularity for the 3D
Euler equations. Indeed, the computations by Boratav and Pelz 
\cite{BP94} and Pelz \cite{Pelz98} indicate a more singular
self-similar type of blow-up. Pelz's computations also fall in the 
critical case of the non-blowup theory of Deng-Hou-Yu \cite{DHY05a,DHY05b}.
We are currently investigating this problem numerically using even higher
resolutions. We will report the results elsewhere.

\vspace{0.2in}
\centerline{\bf \large Appendix. Corrections to Some Misprints in \cite{Kerr93}} 

\vspace{0.1in}
In this appendix, we explain the corrections that we make
regarding the misprints in the description of the initial condition in
\cite{Kerr93}. There are two constraints on the
initial vorticity. The first one is that it must be divergence free.
The second one is that it must satisfy the periodic boundary condition.
It is obvious that we have
\begin{equation}   \nabla \cdot ( \omega_x , \; \omega_y, \; \omega_z ) = 0 .
\end{equation}
Thus the divergence free constraint on the initial vorticity implies
that $\omega ( r )$ must satisfy
\[
\nabla \omega ( r ) \cdot (\omega_x , \;\omega_y ,\; \omega_z ) = 0.
\]

The analytic expression of the initial vorticity profile in \cite{Kerr93}
does not satisfy the above constraint due to a few typos in various formula.
We correct these typos by comparing the analytic formula with the formula
that were actually used by Kerr in his Fortran subroutine that generates
the initial data. Below we would list these typos and point out the
corrections.
\begin{enumerate}
  \item In equation (\ref{trajectory}), the original expression in
   \cite{Kerr93} was written as $[ x_0 + x ( s ), y, z_0 + z ( s ) ]$.
   We remove $x_0$ and $z_0$ from (\ref{trajectory}) since the definition
   of $x(s)$ and $z(s)$ has already taken $x_0$ and $z_0$ into account.
                                                                                           
  \item In (\ref{x-s}), the original expression in \cite{Kerr93} was given
   as $x_0 + \delta_x \cos ( s )$. This would violate the divergence free
   condition. We correct this by replacing $\cos ( s )$ with
   $\cos \left (\pi s/L_x \right )$.
                                                                                           
  \item In (\ref{z-s}), the original expression was $z_0 + \delta_z \cos ( s )$.
   This again violates the divergence free condition. We correct it by replacing    
   $\cos ( s )$ with $\cos \left (\pi s/L_z \right )$.
                                                                                           
  \item In (\ref{w-x}), the last factor in the original expression was
   $\sin ( \pi s ( y ))$. We correct it by replacing $\sin ( \pi s ( y ))$
   with $\sin \left ( \pi s ( y )/L_x \right )$.
                                                                                           
  \item In (\ref{w-z}), the last factor in the original expression was
   $\sin ( \pi s ( y ))$. We correct it by replacing $\sin ( \pi s ( y ))$
   with $\sin \left ( \pi s ( y )/L_z \right )$.
\end{enumerate}
With the above corrections, it can be verified that both the divergence free
condition and the periodic boundary conditions are satisfied.

Based on Kerr's subroutine that generates the initial data, we also make
one minor modification in the definition of $f(r)$. The original equation
in \cite{Kerr93} for (\ref{f}) was given by
\begin{equation}
  f ( r ) = \frac{- r^2}{1 - r^2} + r^2 \left( 1 + r^2 + r^4 \right) .
\end{equation}
After studying Kerr's code, we found that the function $f(r)$ in his
code was actually defined using (\ref{f}) instead of the above formula.
The difference lies in the first factor in the
second term of the above equation. What was used in Kerr's code
was $r^4$ instead $r^2$ in the above equation. This minor modification
has little effect on the behavior of the solution from our computational
experience. We make this minor modification to $f(r)$ in order to match 
exactly the initial condition that was actually used in Kerr's computations.

\vspace{0.2in}
\noindent
{\bf Acknowledgments.}
We would like to thank Prof. Lin-Bo Zhang from the Institute of 
Computational Mathematics in Chinese Academy of Sciences (CAS) for 
providing us with the computing resource to perform this large 
scale computational project. Additional computing resource was 
provided by the Center of High Performance Computing in CAS. We 
also thank Prof. Robert Kerr for providing us with his Fortran 
subroutine that generates his initial data. This work was in part 
supported by NSF under the NSF FRG grant DMS-0353838 and ITR 
Grant ACI-0204932. Part of this work was done while Hou visited 
the Academy of Systems and Mathematical Sciences of CAS in the
summer of 2005 as a member of the Oversea Outstanding Research 
Team for Complex Systems. Finally, we would like to thank Profs. 
Hector Ceniceros and Robert Kerr for their valuable
comments on the original manuscript.

\bibliographystyle{amsplain}
\bibliography{bib}

\providecommand{\bysame}{\leavevmode\hbox to3em{\hrulefill}\thinspace}
\providecommand{\MR}{\relax\ifhmode\unskip\space\fi MR }
\providecommand{\MRhref}[2]{%
  \href{http://www.ams.org/mathscinet-getitem?mr=#1}{#2}
}
\providecommand{\href}[2]{#2}
\begin{thebibliography}{10}

\bibitem{AG89}
C.~Anderson and C.~Greengard, \emph{The vortex ring merger problem at infinite
  reynolds number}, Comm. Pure Appl. Maths \textbf{42} (1989), 1123.

\bibitem{BKM84}
J.~T. Beale, T.~Kato, and A.~Majda, \emph{Remarks on the breakdown of smooth
  solutions of the 3-{D} {E}uler equations}, Comm. Math. Phys. \textbf{96}
  (1984), 61--66.

\bibitem{BP94}
O.~N. Boratav and R.~B. Pelz, \emph{Direct numerical simulation of transition
  to turbulence from a high-symmetry initial condition}, Phys. Fluids
  \textbf{6} (1994), no.~8, 2757--2784.

\bibitem{BPZ92}
O.~N. Boratav, R.~B. Pelz, and N.~J. Zabusky, \emph{Reconnection in
  orthogonally interacting vortex tubes: Direct numerical simulations and
  quantifications}, Phys. Fluids A \textbf{4} (1992), no.~3, 581--605.

\bibitem{Caf93}
R.~Caflisch, \emph{Singularity formation for complex solutions of the 3{D}
  incompressible {E}uler equations}, Physica D \textbf{67} (1993), 1--18.

\bibitem{Chorin82}
A.~Chorin, \emph{The evolution of a turbulent vortex}, Commun. Math. Phys.
  \textbf{83} (1982), 517.

\bibitem{Const94}
P.~Constantin, \emph{Geometric statistics in turbulence}, SIAM Review
  \textbf{36} (1994), 73.

\bibitem{CFM96}
P.~Constantin, C.~Fefferman, and A.~Majda, \emph{Geometric constraints on
  potentially singular solutions for the 3-{D} {E}uler equation}, Commun. in
  PDEs. \textbf{21} (1996), 559--571.

\bibitem{DHY05a}
J.~Deng, T.~Y. Hou, and X.~Yu, \emph{Geometric properties and non-blowup of
  3-{D} incompressible {E}uler flow}, Comm. in PDEs. \textbf{30} (2005), no.~1,
  225--243.

\bibitem{DHY05b}
\bysame, \emph{Improved geometric conditions for non-blowup of 3{D}
  incompressible {E}uler equation}, accepted by Comm. in PDEs. (2005).

\bibitem{EFM70}
D.~G. Ebin, A.~E. Fischer, and J.E. Marsden, \emph{Diffeomorphism groups,
  hydrodynamics and relativity}, The 13th Biennial Seminar of Canadian
  Mathematical Congress (J.~Vanstone, ed.), 1970, pp.~135--279.

\bibitem{GMG98}
R.~Grauer, C.~Marliani, and K.~Germaschewski, \emph{Adaptive mesh refinement
  for singular solutions of the incompressible {E}uler equations}, Phys. Rev.
  Lett. \textbf{80} (1998), 19.

\bibitem{GS91}
R.~Grauer and T.~Sideris, \emph{Numerical computation of three dimensional
  incompressible ideal fluids with swirl}, Phys. Rev. Lett. \textbf{67} (1991),
  3511.

\bibitem{Kerr93}
R.~M. Kerr, \emph{Evidence for a singularity of the three dimensional,
  incompressible {E}uler equations}, Phys. Fluids \textbf{5} (1993), no.~7,
  1725--1746.

\bibitem{Kerr97}
\bysame, \emph{{E}uler singularities and turbulence}, 19th ICTAM Kyoto '96
  (T.~Tatsumi, E.~Watanabe, and T.~Kambe, eds.), Elsevier Science, 1997,
  pp.~57--70.

\bibitem{Kerr99}
\bysame, \emph{The outer regions in singular {E}uler}, Fundamental problematic
  issues in turbulence (Birkh\"auser) (Tsnober and Gyr, eds.), 1999.

\bibitem{Kerr04}
\bysame, \emph{Velocity and scaling of collapsing {E}uler vortices}, Phys.
  Fluids (to appear).

\bibitem{KH89}
R.~M. Kerr and F.~Hussain, \emph{Simulation of vortex reconnection}, Physica D
  \textbf{37} (1989), 474.

\bibitem{MB02}
A.~J. Majda and A.~L. Bertozzi, \emph{Vorticity and {I}ncompressible {F}low},
  Cambridge University Press, 2002.

\bibitem{MH89}
M.~V. Melander and F.~Hussain, \emph{Cross linking of two antiparallel vortex
  tubes}, Phys. Fluids A (1989), 633--636.

\bibitem{Pelz98}
R.~B. Pelz, \emph{Locally self-similar, finite-time collapse in a high-symmetry
  vortex filament model}, Phys. Rev. E \textbf{55} (1997), no.~2, 1617--1626.

\bibitem{PS90}
A.~Pumir and E.~E. Siggia, \emph{Collapsing solutions to the 3-{D} {E}uler
  equations}, Phys. Fluids A \textbf{2} (1990), 220--241.

\bibitem{SMO93}
M.~J. Shelley, D.~I. Meiron, and S.~A. Orszag, \emph{Dynamical aspects of
  vortex reconnection of perturbed anti-parallel vortex tubes}, J. Fluid Mech.
  \textbf{246} (1993), 613--652.

\bibitem{Chorin06}
P.~Stinis and A.~Chorin, \emph{Numerical scaling analysis of the small-scale
  structure in turbulence}, LBNL report LBNL-59490, Mathematics Dept. (2006).

\end{thebibliography}

\end{document}